\documentclass[reprint, amsmath,amssymb, nofootinbib, aps]{revtex4-2}

\usepackage{tikz}
\usepackage{color}
\usepackage{ulem}
\usepackage{subfigure}
\usepackage{graphicx}
\usepackage{dcolumn}
\usepackage{bm}
\usepackage[colorlinks, linkcolor=blue, anchorcolor=blue, citecolor=blue, urlcolor=blue]{hyperref}
\usepackage{url}
\usepackage{multirow}
\usepackage{booktabs}
\usepackage{subfigure}
\definecolor{lime}{HTML}{A6CE39}
\DeclareRobustCommand{\orcidicon}{%
	\begin{tikzpicture}
	\draw[lime, fill=lime] (0,0)
	circle [radius=0.16]
	node[white] {{\fontfamily{qag}\selectfont \tiny ID}};
	\draw[white, fill=white] (-0.0625,0.095)
	circle [radius=0.007];
	\end{tikzpicture}
	\hspace{-2mm}
}

\foreach \x in {A, ..., Z}{%
	\expandafter\xdef\csname orcid\x\endcsname{\noexpand\href{https://orcid.org/\csname orcidauthor\x\endcsname}{\noexpand\orcidicon}}
}

\begin{document}

\title{Nonreciprocity enhanced Quantum Gyroscopes based on Surface Acoustic Waves}
\author{Yuting Zhu\orcidA{}$^{1,2,3}$}
\author{Shibei Xue\orcidB{}$^{1,2,3}$}
\email{shbxue@sjtu.edu.cn}
\author{Fangfang Ju$^{4}$}
\author{Haidong Yuan\orcidC{}$^{5}$}

\affiliation{$^{1}$School of Automation and Intelligent Sensing, Shanghai Jiao Tong University, Shanghai, 200240, P. R. China}
\affiliation{$^{2}$Key Laboratory of System Control and Information Processing, Ministry of Education of China, Shanghai, 200240, P. R. China}
\affiliation{$^{3}$Shanghai Key Laboratory of Perception and Control in Industrial Network Systems, Shanghai,  200240, P. R. China}
\affiliation{$^{4}$School of Physics and Electronics, Hunan Normal University, Changsha, P. R. China}
\affiliation{$^{5}$Department of Mechanical and Automation Engineering, The Chinese University of Hong Kong, Hong Kong SAR, P. R. China}

\date{\today}

\begin{abstract}
{Surface acoustic waves (SAWs), as Rayleigh waves generated by elastic media, have been used in gyroscopes for over 40 years due to their unique propagation characteristics. However, their working principle, based on Coriolis effects, has become increasingly ineffective for addressing modern sensing challenges in complex scenarios. Fortunately, recent advancements in quantized SAWs offer a promising solution: SAWs operating at extremely low pump powers (approximately at the single-phonon level) can exhibit substantial quantum coherence, enabling investigations into the fundamental limits of SAW gyroscopes as constrained by the Heisenberg uncertainty relation. In particular, when multiple SAWs couple to a common waveguide at distinct locations, the nonlocality arising from the spatial separation among coupling points induces directional coupling between the SAWs. To elucidate this directionality, we propose a quantum gyroscope characterized by multiple-point couplings. Unlike traditional single-point coupling designs, our gyroscope exhibits distinctive time-delayed dynamics that depend on the system's topologies. We emphasize that these dynamics invalidate the Markovian approximation, even when the time delay is relatively small. Through a comprehensive analysis of all possible topologies, we observe that the directional coupling implies an inherent nonreciprocal transfer. This nonreciprocity confers significant advantages to our gyroscope compared to traditional designs, notably enhancing both the signal-to-noise ratio  and sensitivity. Specifically, it enables the extraction of output signals that would otherwise be obscured by noise. Consequently, our findings suggest that systems with multiple-point couplings and the associated nonreciprocity can serve as valuable resources for advancing quantum sensing technologies.}

\end{abstract}
\maketitle
\section{Introduction\label{Sec1}}
{Surface acoustic waves (SAWs) are mechanical waves generated by elastic media via the piezoelectric effect, whose energy is concentrated near the surface or interface of the media and propagates along that surface \cite{Rayleigh1885, Friend2011}. The amplitude of SAWs decays exponentially with depth into the material, typically diminishing within one or two wavelengths. This distinctive propagation characteristic makes SAWs an ideal platform for sensing applications, particularly in gyroscopes used to measure angular velocity \cite{Oh2015, Kukaev2025}. Due to not having any moving parts and vulnerable suspensions, SAW gyroscopes offer better vibration and shock resistance compared to traditional gyroscopes, such as rotator gyroscopes \cite{Barr1961, Craig1972} and MEMS gyroscopes \cite{Acar2008, Armenise2010}. Additionally, the straightforward fabrication process of SAW gyroscopes provides a competitive mass-production price \cite{Kukaev2025}. However, these advantages do not imply that SAW gyroscopes are without limitations. Their working principle makes them increasingly challenging to meet the ultra-high sensitivity demand in complex scenarios, e.g., high-speed aircraft flying around the Earth. SAW gyroscopes rely on the Coriolis effect in a rotating frame \cite{Kukaev2025}, which requires not only high-power SAWs but also dominant Coriolis forces to suppress centrifugal effects. In many modern sensing applications, the Coriolis force does not always dominate. Therefore, improving SAW gyroscopes to overcome these challenges remains an open question.}

\begin{figure*} [bt]
\centering
\subfigure{\includegraphics[width=0.46\textwidth]{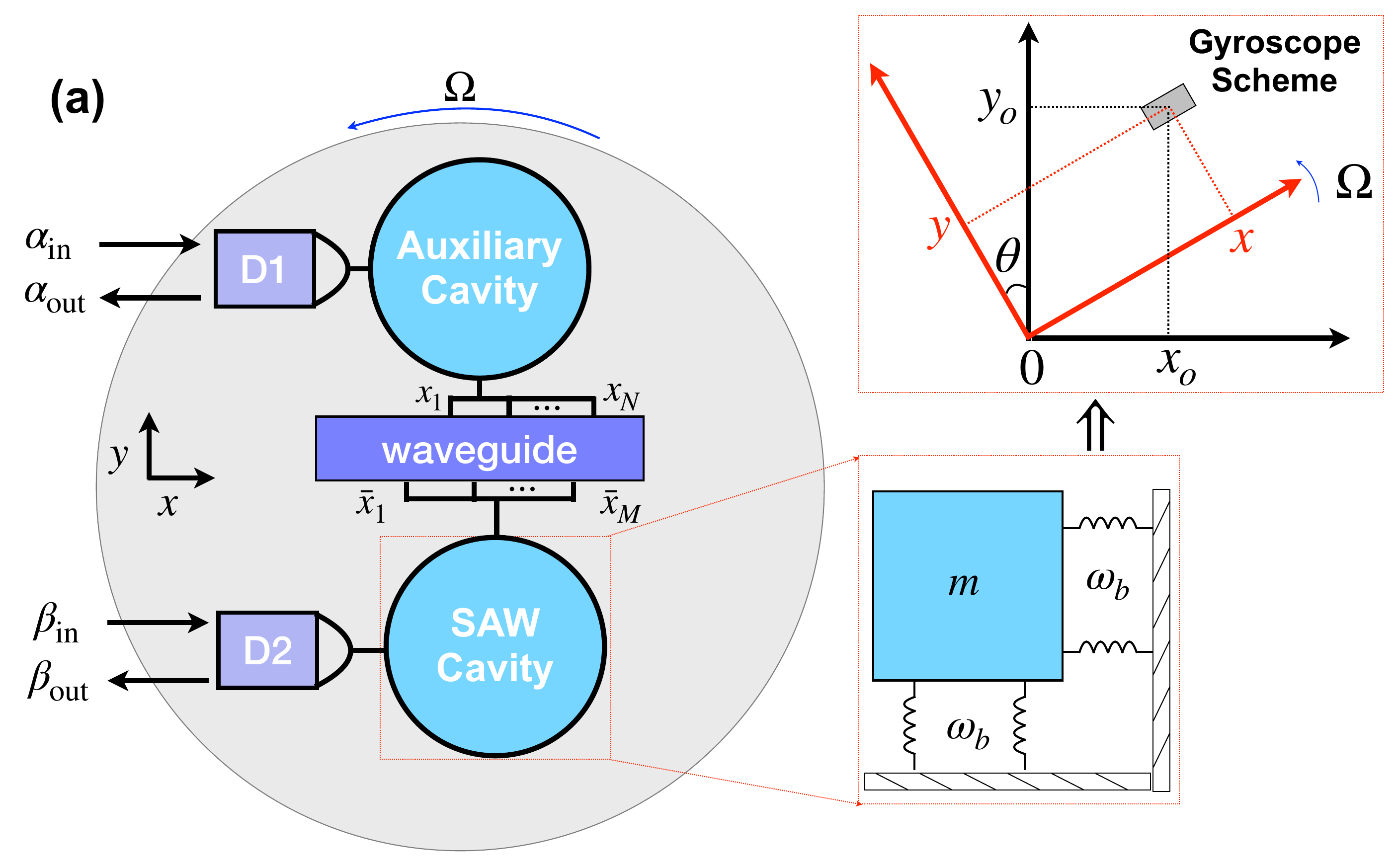}}
\subfigure{\includegraphics[width=0.46\textwidth]{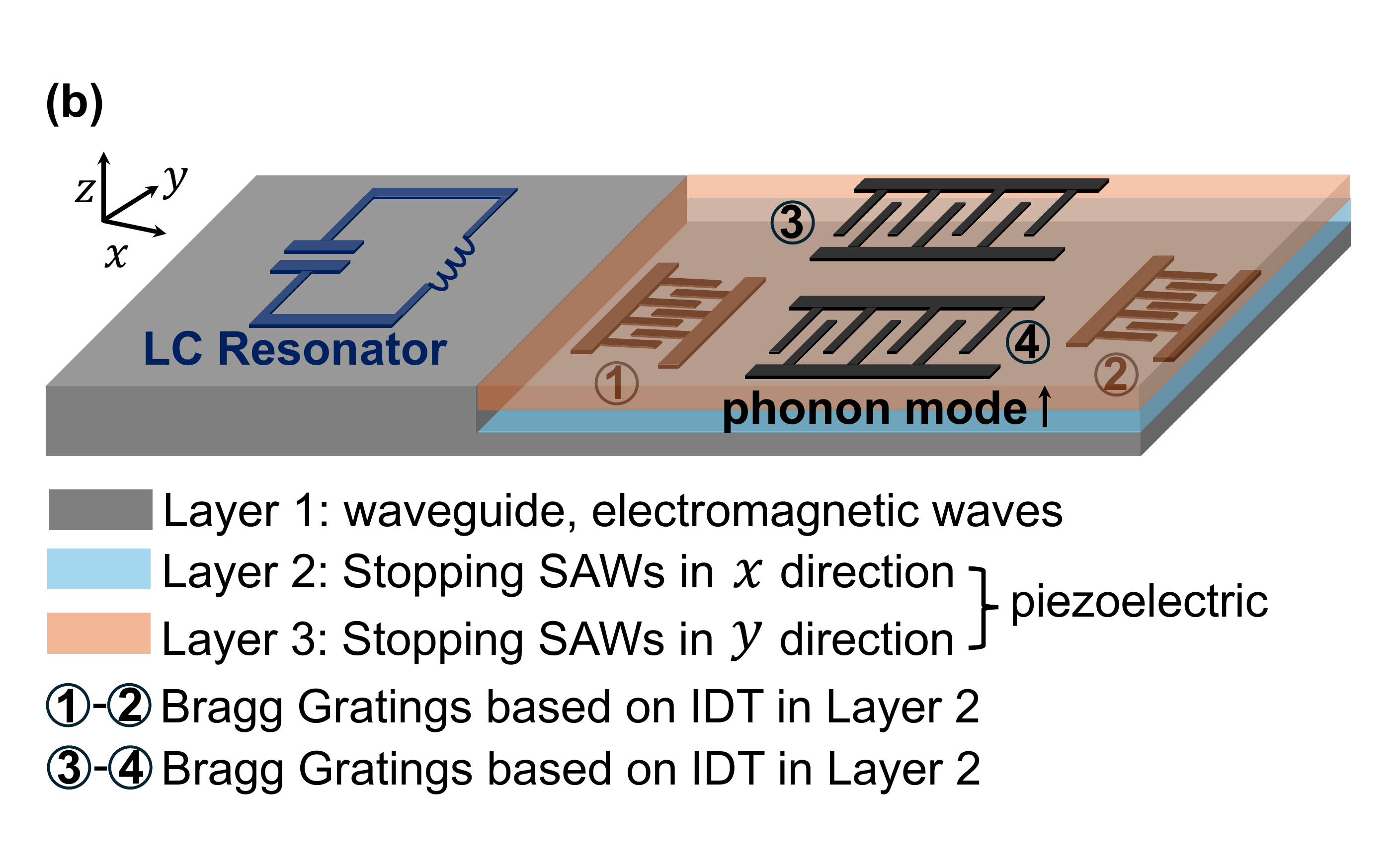}}
\caption{(Color online) Schematic of the quantum gyroscope. (a) Toy model. The SAW cavity supports two orthogonal modes $x$ and $y$, and they couple together when the plate rotates at an unknown angular velocity $\Omega$. Consequently, the SAW cavity serves as a fundamental gyroscope configuration in the $x-y$ plane. By coupling the $x$ mode and the auxiliary cavity to the same waveguide at distant points, we establish a coupled quantum giant-cavity system. To analyze the underlying directional coupling, two detection ports, D1 and D2, are employed for readout. Port D1 connects to the auxiliary cavity, while port D2 couples to the $x$ mode.  {(b) An accessible realization in experiments (without detectors). The whole setup is fabricated by multiple-layer  lithograph technology. The layer 1 consists of a sapphire substrate whose surface is covered with an aluminum membrane. The aluminum membrane is used for etching the LC resonator (auxiliary cavity) and the waveguide, where the multiple-point couplings between them is realized by multiple capcitors (not given).  The layers 2 and 3 consist of elastic media exhibiting the piezoelectric effect, used to generate SAWs in the $x$ and $y$ directions, respectively. Additionally, one can etch interdigital transducers (IDTs) in layers 2 and 3 to form Bragg gratings, which truncate the traveling SAWs into standing waves, thereby constructing a double-mode SAW cavity. Besides, one can also use multiple IDTs to construct the muliplt-point coupling between $x$ mode and the waveguide (not given).}}
\label{Model}
\end{figure*}

In addressing the question above, the recent study \cite{Zhu2024} offers a potential solution, wherein the authors utilize quantized SAWs to construct an optomechanical-like gyroscope \cite{Davuluri2016, Davuluri2017, Li2018A, Li2018B, Li2018C, Lavrik2019}. {Different from classic SAWs with high power, quantized SAWs with extremly low power (about pW-nW magnitude) do not transport electrons or holes in the semiconductor; instead, they transport phonons \cite{Gustafsson2014}. }This characteristic endows quantized SAWs with remarkable quantum coherence \cite{Schuetz2015, Chu2017, Manenti2017, Chu2018, Satzinger2018, Andersson2019, Ekstrom2019, Kannan2020, Andersson2020, Vadiraj2021}, enabling exploration of the quantum limit of sensitivity permitted by the Heisenberg uncertainty relation \cite{Clerk2010, Aspelmeyer2014, Zhu2024}. In particular, when multiple SAWs couple to the same waveguide at distant points, forming a so-called coupled quantum giant system \cite{Kockum2018, Kannan2020, Feng2021, Zhu2022B}, the inherent nonlocality arising from the separation of coupling points induces unique time-delayed dynamics, rendering the Markovian approximation inadequate \cite{Zhu2022A, Zhu2022B}. Consequently, these time-delayed dynamics give rise to directional coupling among SAWs \cite{Zhu2022B}. This directional coupling implies nonreciprocal transfer---that is, unequal transfer when reversing the source and receiver \cite{Deak2012, Lau2018}---and depends on both the system topology and the number of coupling points. {However, existing studies \cite{Kockum2018, Kannan2020, Feng2021, Zhu2022B} have focused exclusively on specific topologies with two coupling points and have relied on the Markovian approximation. This approach not only neglects other possible topologies with a greater number of coupling points but also leaves the underlying nonreciprocity unexplored}. Importantly, this nonreciprocity could serve as a powerful resource for enhancing sensing capabilities \cite{Lau2018}. Therefore, a detailed and comprehensive analysis of the nonreciprocity induced by different topologies, as well as its impact on sensing performance, remains an open problem in coupled quantum giant systems

To address these challenges, we propose a quantum gyroscope based on two coupled giant cavities, where the SAW cavity serves as the fundamental gyroscope frame in the $x-y$ plane. {Unlike previous investigations that restricted the number of coupling points to two \cite{Kockum2018, Kannan2020, Zhu2022B}, we systematically consider all possible topologies with a greater number of coupling points and thoroughly analyze the resulting time-delayed dynamics.} We further find that the dynamics in multiple-point-coupling systems can exhibit pronounced non-Markovian effects, rendering the Markovian approximation inapplicable. We then comprehensively explore the underlying nonreciprocity across various topologies and find that this nonreciprocity is an intrinsic property of multiple-point-coupling structures, guaranteed by the system's causality. Furthermore, we conduct a detailed analysis of the impact of nonreciprocity on sensing performance, focusing on both the SNR and sensitivity. Our results show that both indices can be significantly enhanced due to nonreciprocity. Compared to traditional single-point-coupling structures, the improved signal-to-noise ratio (SNR) in our multiple-point-coupling system not only renders previously noise-overwhelmed output signals readable but also enhances sensitivity under identical parameters. This indicates that the multiple-point-coupling system and the associated nonreciprocity it entails can serve as a powerful resource for future quantum sensing applications.

The remainder of this paper is organized as follows: In Sec. \ref{Sec2}, we present the model of the gyroscope and conduct a comprehensive analysis of its time-delayed dynamics. Following a frequency analysis, we discuss the transfer function and nonreciprocity in Sec. \ref{Sec3} and Sec. \ref{Sec4}, respectively. Discussions on sensing performances are provided in Sec. \ref{Sec5}, and we compare our gyroscope with traditional ones in Sec. \ref{Sec6}. To guide experiments, we also provide discussions on experimental feasibility of our gyroscope in Sec. \ref{Sec7}. Finally, we conclude our work in Sec. \ref{Sec8}.

\section{Model of the gyroscope and its time-delayed dynamics\label{Sec2}}

The gyroscope consists of double-coupled cavities mounted on a rotating plate, where the SAW cavity functions as a double-mode mechanical oscillator with an effective mass $m$ and frequency $\omega_b$, as illustrated in Fig. \ref{Model} (a). When the plate rotates, its angular velocity $\Omega$ becomes the coupling strength between the orthogonal modes $x$ and $y$ of the SAW cavity. Consequently, the SAW cavity serves as the fundamental framework for gyroscopes in the $x-y$ plane. The coordinate system applies to the entire setup, and our objective is to measure the unknown angular velocity $\Omega$ using external probes. Unlike traditional coupled-cavity gyroscopes \cite{Zhu2024, Davuluri2016, Davuluri2017, Li2018, Lavrik2019}, the coupling between the auxiliary cavity and the $x$ mode is achieved by connecting them to a common waveguide at distinct points, thereby forming a so-called quantum giant system \cite{Kockum2014, Guo2017, Kockum2018, Andersson2019, Guo2020, Kannan2020, Du2021A, Du2021B, Cai2021, Feng2021, Zhu2022A, Zhu2022B, Peng2023, Zhang2025}. Recent advancements \cite{Zhu2022B} indicate that this multiple-point coupling can exhibit directionality if the coupling points follow a specific topology. To thoroughly investigate this underlying directionality, we have established two detection ports, D1 and D2, for readout: port D1 connects to the auxiliary cavity, while port D2 couples to the $x$ mode. {In experiments, the setup can be fabricated using multilayer photolithography technology, as shown in Fig. \ref{Model}(b). Layer 1 is used to etch the auxiliary cavity and the waveguide and to transport the electromagnetic field. Layers 2 and 3 consist of elastic media exhibiting the piezoelectric effect, used to generate SAWs in the $x$ and $y$ directions, respectively. }

According to the model, the setup has the Hamiltonian
\begin{eqnarray}
H&=&H_{gyro}+H_g \label{Hamiltonian}
\end{eqnarray}
with
\begin{subequations}
\begin{eqnarray}
H_{gyro}&=&\omega_a a^\dag a+\omega_b (b_x^\dag b_x+ b_y^\dag b_y)+i\Omega(b_x^\dag b_y-H.c.)\nonumber\\
&&-i(\sqrt{\kappa_a}a^\dag \alpha_{\mathrm{in}}e^{-i\omega_d t}-H.c.)\nonumber\\
&&-i(\sqrt{\kappa_b}b_x^\dag \beta_{\mathrm{in}}e^{-i\omega_d t}-H.c.),\\
H_g&=&\int\mathrm{d}\omega~\omega~c_\omega^\dag c_\omega\nonumber\\
&&-i\sum_{n=1}^N\sqrt{\frac{\gamma_n}{2\pi}}\int\mathrm{d}\omega (a^\dag c_\omega e^{i\omega \tau_n}-H.c.)\nonumber\\
&&-i\sum_{m=1}^M\sqrt{\frac{\bar{\gamma}_m}{2\pi}}\int\mathrm{d}\omega (b_x^\dag c_\omega e^{i\omega \bar{\tau}_m}-H.c.).\label{GiantCavityCoupling}
\end{eqnarray}
\end{subequations}
The Hamiltonian $H_{gyro}$ denotes the basic Hamiltonian of the gyroscope, consistent with traditional designs \cite{Li2018A, Li2018B, Zhu2024}. In this context, the auxiliary cavity, the $x$ mode, and the $y$ mode are denoted by the bosonic annihilation operators $a$, $b_x$, and $b_y$, respectively. The probe field $\alpha_{\mathrm{in}}e^{-i\omega_d t}$ ($\beta_{\mathrm{in}}e^{-i\omega_d t}$) enters the auxiliary cavity $a$ (the $x$ mode) from the port D1 (D2) with a strength $\kappa _a$ ($\kappa_b$). The Hamiltonian $H_g$ describes the position-dependent interaction between the auxiliary cavity and the $x$ mode, achieved by coupling them to a common waveguide represented by the annihilation operator $c_\omega$. The corresponding coupling strengths between the cavity and the waveguide are $\gamma_n$ and $\bar{\gamma}_m$, respectively. Here, the positions $x_n$ and $\bar{x}_m$ are expressed as time delays $\tau_n=x_n/v_g>0$ and $\bar{\tau}_m=\bar{x}_m/v_g>0$ with group velocity $v_g$. For simplicity, we consider $\gamma_n=\bar{\gamma}_m=\gamma$ in following contents.

Without loss of generality, the mode $x$ ($y$) is assumed to be influenced by thermal noise $f_{x}$ ($f_y$) with decay rate $\gamma_{x}$ ($\gamma_{y}$). Under a rotating frame with respect to the driving frequency $\omega_d$, the system satisfies the following time-delayed Heisenberg-Langevin equation
\begin{equation}
\begin{aligned}
\dot{\boldsymbol{\mathrm{x}}}(t)&=A_o\boldsymbol{\mathrm{x}}(t)-B_o\boldsymbol{\mathrm{u}}_{\mathrm{in}}(t)\\
&+{L}(t;\tau_n,\bar{\tau}_m)-{L}_{\mathrm{in}}(t;\tau_n,\bar{\tau}_m).
\end{aligned}
\label{Eom}
\end{equation}
{To avoid excessive mathematical detail in the main text, we have placed its full derivation and detailed expressions in Appendix \ref{AppendixA} for easier reading. The first line of Eq. \eqref{Eom} describes time-local dynamics with the mode vector $\boldsymbol{\mathrm{x}}=({a}, {b}_x, {b}_y)^T$ and the input vector $\boldsymbol{\mathrm{u}}_{\mathrm{in}}(t)=(\alpha_{\mathrm{in}}(t), \beta_{\mathrm{in}}(t), f_x(t), f_y(t))^T$, while the second line represents time-nonlocal dynamics. Notably, this nonlocal component lies at the core of the considered multiple-point coupling system, as it not only reflects a characteristic non-Markovian time-delayed effect but also implies an underlying directional coupling.} It takes the form
\begin{subequations}
\begin{eqnarray}
&&{L}(t;\tau_n,\bar{\tau}_m)=A_{f}\boldsymbol{\mathrm{x}}_{f}(t;\tau_n,\bar{\tau}_m)+A_{c}\boldsymbol{\mathrm{x}}_{c}(t;\tau_n,\bar{\tau}_m),
\label{UniqueDynamicsGiantMode}\\
&&{L}_{\mathrm{in}}(t;\tau_n,\bar{\tau}_m)=B_c\boldsymbol{\mathrm{c}}_{\mathrm{in}}(t;\tau_n,\bar{\tau}_m).\label{UniqueDynamicsGiantIn}
\end{eqnarray}
\end{subequations}
Here, the vectors
\begin{subequations}
\begin{eqnarray}
&&\boldsymbol{\mathrm{x}}_{f}(t;\tau_n,\bar{\tau}_m)=\bigg(a(t;\tau_n, \tau_{n'}), b_x(t;\bar{\tau}_m, \bar{\tau}_{m'}), b_y(t)\bigg)^T\\
&&\boldsymbol{\mathrm{x}}_{c}(t;\tau_n,\bar{\tau}_m)=\bigg(a(t;\bar{\tau}_{m}, \tau_n), b_x(t;\tau_n, \bar{\tau}_{m}), b_y(t)\bigg)^T\\
&&\boldsymbol{\mathrm{c}}_{\mathrm{in}}(t;\tau_n,\bar{\tau}_m)=\bigg(c_{\mathrm{in}}(t-\tau_n), c_{\mathrm{in}}(t-\bar{\tau}_m)\bigg)^T
\end{eqnarray}
\end{subequations}
with the operator $O(t; h_i, s_j)=O\big(t-(h_i-s_j)\big)$ and the quantum noise input  $c_{\mathrm{in}}(t)=\frac{1}{\sqrt{2\pi}}\int\mathrm{d}\omega~c_\omega(0)e^{-i\omega t}$ led by the waveguide, describe the time-delayed dynamics of the system. The coefficient matrices
\begin{subequations}
\begin{eqnarray}
&&A_{f}=\mathrm{diag}(
\Gamma(\omega_d; \tau_n,\tau_{n'}),\Gamma(\omega_d; \tau_m,\tau_{m'}),0),\\
&&A_{c}=
\begin{pmatrix}
0 & \Gamma(\omega_d; \tau_n,\bar{\tau}_{m}) & 0	\\
\Gamma(\omega_d; \bar{\tau}_m,\tau_{n}) & 0 & 0 \\
0 & 0 & 0 \\
\end{pmatrix},\label{GiantCavityA}\\
&&B_c=
\begin{pmatrix}
 \sum_{n=1}^N\sqrt{\gamma}e^{i\omega_d \tau_n} & 0  \\
0 & \sum_{m=1}^M\sqrt{\gamma}e^{i\omega_d\bar{\tau}_m}  \\
0 & 0
\end{pmatrix},
\end{eqnarray}
\end{subequations}
characterized by the effective decay rate $\Gamma(\omega_d;h_i, s_j)=-\gamma\sum_i\sum_j\Theta(h_i-s_j)e^{i\omega_d(h_i-s_j)}$ and the Heaviside step function
\begin{equation}
\Theta(h_i-s_j)=
\begin{cases}
1, & h_i-s_j>0\\
\frac{1}{2}, & h_i-s_j=0\\
0, & h_i-s_j<0
\end{cases},
\end{equation}
indicate the directional coupling between the auxiliary cavity and the $x$ mode.

For the Heaviside step function in Eq. \eqref{GiantCavityA}, there are two important points to note. (i) It takes different forms in the nondiagonal terms of Eq. \eqref{GiantCavityA}, such that its cutoff property renders the coupling between the modes $a$ and $b_x$ unidirectional for each pair of unequal coupling points $\tau_n \neq \bar{\tau}_m$. This implies that the mode at the former position can influence the latter, but not vice versa. We emphasize that this directional coupling, resulting from time delays, is guaranteed by the causality of the system: when the $a$ mode is at the former position, i.e., $\tau_n < \bar{\tau}_m$, the dynamics of the $b_x$ mode always lag behind the $a$ mode, such that the state of the $a$ mode unidirectionally transfers to the $b_x$ mode. (ii) Ignoring the Heaviside step functions could lead to a misuse of the Markovian approximation. In the Markovian approximation, the time-delayed terms are approximated as $\Gamma(\omega_d;h_i, s_j)O(t; h_i, s_j)\approx-\gamma\sum_{i}\sum_{j}e^{i\omega_d|h_i-s_j|}O(t)$. Clearly, the Markovian approximation allows both $b_x(t-(\tau_n-\bar{\tau}_m))$ and $a(t-(\bar{\tau}_m-\tau_n))$ to play roles, indicating that the $a \leftrightarrow b_x$ coupling is bidirectional, regardless of which mode is at the former position. However, Eq. \eqref{GiantCavityA} demonstrates that bidirectional coupling only occurs under specific conditions, such as when the condition $\tau_n=\bar{\tau}_m$ strictly holds. Strictly speaking, the directional coupling is essentially a non-Markovian effect, i.e., nonlocal evolution in the time domain, while the Markovian approximation is a local transformation that manually eliminates this nonlocality \cite{Zhu2022A}. Therefore, the Markovian approximation does not apply to multiple-point-coupling systems, even in the limit Markovian regime $\tau_{nm}=|\tau_n-\bar{\tau}_m|\rightarrow0$. In this regime, the relative time delay $\tau_{nm}$ can be negligible, but this does not imply that the coupling points $\tau_n$ and $\bar{\tau}_m$ strictly overlap; thus, the coupling remains unidirectional due to the cutoff of the Heaviside step functions.

Besides the equations of motion \eqref{Eom}, the Heisenberg-Langevin equation also gives us the input-output relation
\begin{eqnarray}
\begin{pmatrix}
\alpha_{\mathrm{out}}(t) \\
\beta_{\mathrm{out}}(t)
\end{pmatrix}
=\begin{pmatrix}
\alpha_{\mathrm{in}}(t) \\
\beta_{\mathrm{in}}(t)
\end{pmatrix}
+\begin{pmatrix}
\sqrt{\kappa_a} & 0 \\
0 & \sqrt{\kappa_b}
\end{pmatrix}
\begin{pmatrix}
a(t) \\
b_x(t)
\end{pmatrix}.
\label{InputOutputRelation}
\end{eqnarray}
Notably, it describes the relationship between the two ports, indicating that the time-delayed effect is only reflected in the system's modes $a$ and $b_x$.

\section{Solutions of transfer function \label{Sec3}}
Now, we focus on the solutions of the time-delayed equations of motion \eqref{Eom}. By utilizing the symmetric Fourier transform of an arbitrary operator $O$
\begin{subequations}
\begin{eqnarray}
O(\omega)&=&\int\mathrm{d}t ~O(t)e^{-i\omega t}\label{FourierO},\\
O^\dag(\omega)&=&\int\mathrm{d}t ~O^\dag(t)e^{i\omega t}=[O(\omega)]^\dag\label{FourierOdag},
\end{eqnarray}
\end{subequations}
we can reorganize the solution of Eq. \eqref{Eom} and the input-output relation in a simple form
\begin{eqnarray}
&&\tilde{\boldsymbol{\mathrm{x}}}(\omega)=\mathcal{A}^{-1}(\omega)\mathcal{B}(\omega)\tilde{\boldsymbol{\mathrm{u}}}_{\mathrm{in}}(\omega),\\
\label{Solutions}
&&\boldsymbol{\mathrm{u}}_{\mathrm{out}}(\omega)=G(\omega) \tilde{\boldsymbol{\mathrm{u}}}_{\mathrm{in}}(\omega).
\end{eqnarray}
Here, we have incorporated the $b_y$ mode into the $b_x$ mode, as we are not interested in the dynamics of the $b_y$ mode. As a result, the mode and input vectors become as $\boldsymbol{\mathrm{x}}(\omega)\rightarrow\tilde{\boldsymbol{\mathrm{x}}}(\omega)=(a(\omega), b_x(\omega))^T$ and $\boldsymbol{\mathrm{u}}_{\mathrm{in}}(\omega)\rightarrow\tilde{\boldsymbol{\mathrm{u}}}_{\mathrm{in}}(\omega)=(
\alpha_{\mathrm{in}}(\omega), \beta_{\mathrm{in}}(\omega),  c_{\mathrm{in}}(\omega),f_x(\omega),  f_y(\omega))^T$, and the output vector reads  $\boldsymbol{\mathrm{u}}_{\mathrm{out}}(\omega)=(\alpha_{\mathrm{out}}(\omega), \beta_{\mathrm{out}}(\omega))^T$.For clarity, the detailed forms of the matrices $\mathcal{A}(\omega)$ and $\mathcal{B}(\omega)$ are provided in Appendix \ref{AppendixA}, and the transfer function reads
\begin{eqnarray}
G(\omega)=
\begin{pmatrix}
I_{2\times2} & 0_{2\times3}
\end{pmatrix}
+
\begin{pmatrix}
\sqrt{\kappa_a} & 0  \\
0 & \sqrt{\kappa_b}
\end{pmatrix}\mathcal{A}^{-1}(\omega)\mathcal{B}(\omega).\label{TransferFunction}
\end{eqnarray}
Notably, the matrices $\mathcal{A}(\omega)$ and $\mathcal{B}(\omega)$ satisfy the relation
\begin{eqnarray}
\mathcal{A}(\omega)+\mathcal{A}^\dag(\omega)+\mathcal{B}(\omega)\mathcal{B}^\dag(\omega)\equiv0,\label{ABRelation}
\end{eqnarray}
which further makes the transfer function $G(\omega)$ satisfy the condition
\begin{eqnarray}
|G(\omega)|^2=G(\omega)G^\dag(\omega)\equiv I. \label{EnergyConservation}
\end{eqnarray}
It indicates the energy conservation of the system under gainless situations. In Ref. \cite{Zhang2022}, the authors proved that the condition \eqref{EnergyConservation} applies to all linear gainless quantum systems with single-point couplings, provided the system satisfies condition \eqref{ABRelation}. Our results show that these two conditions also hold for quantum giant systems with multilple-point couplings. In fact, the conditions \eqref{ABRelation} and \eqref{EnergyConservation} naturally arise if the equations of motion are strictly derived from the Heisenberg-Langevin equation\footnote{In Ref. \cite{Zhang2022}, the equations of motion are reformulated as quantum stochastic differential equations (QSDEs), where the integro-differential rule follows the Ito rule. In the standard QSDE framework \cite{Combes2017}, the system must evolve locally in time, implying the absence of non-Markovian memory effects. Therefore, QSDE does not strictly apply to quantum giant systems.}.

\section{Topologies and the underlying Nonreciprocity \label{Sec4}}
\begin{figure} [bt]
\centering
\includegraphics[width=0.48\textwidth]{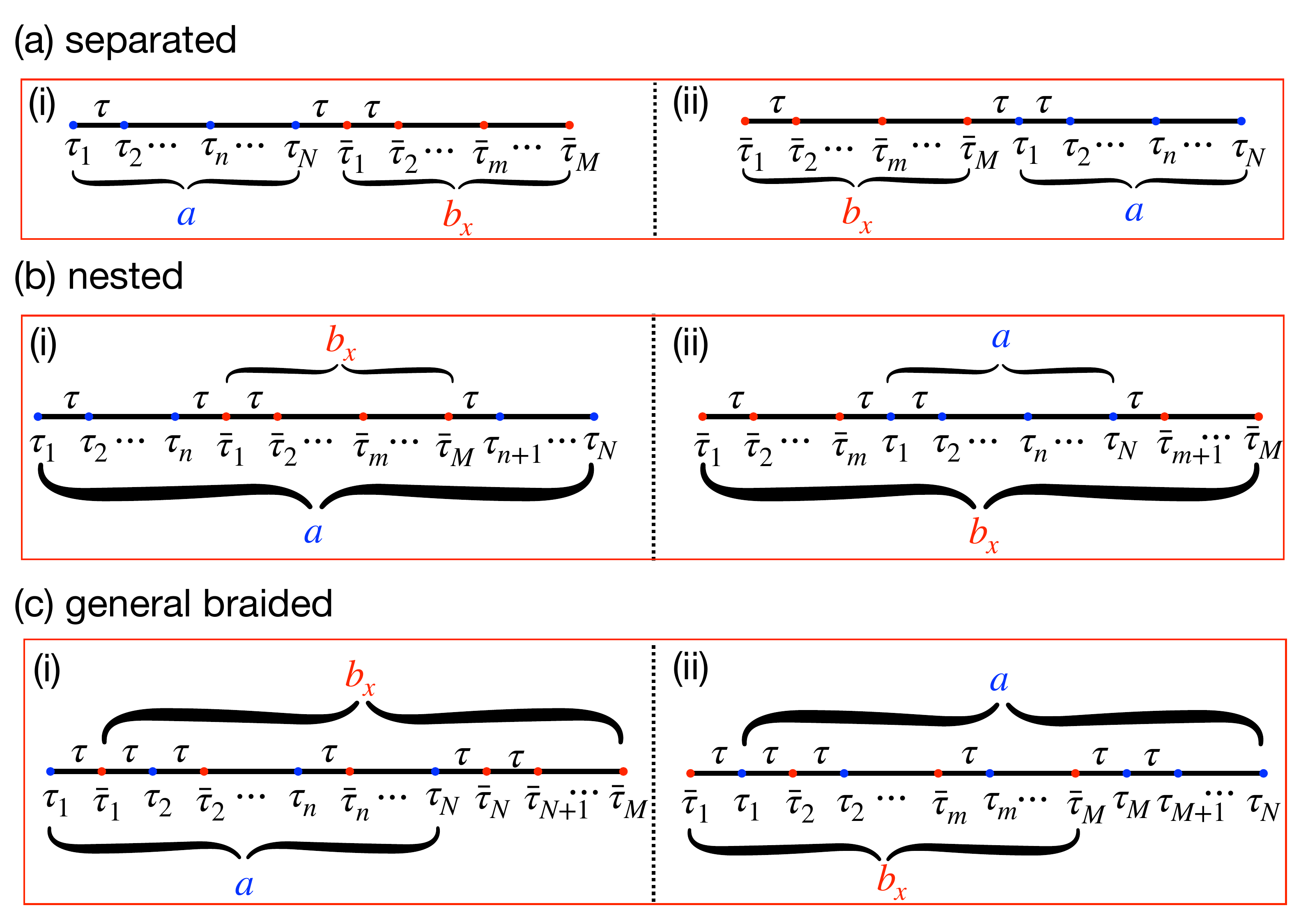}
\caption{(Color online) Schematic of the topologies between modes $a$ and $b_x$. For the double-body system examined in this paper, there are three types of topologies: (a) separated, (b) nested, and (c) braided. The topologies labeled (ii) are mirror images of those labeled (i). }
\label{topology}
\end{figure}
\begin{figure*}[bt]
\subfigure{\includegraphics[width=0.49\textwidth]{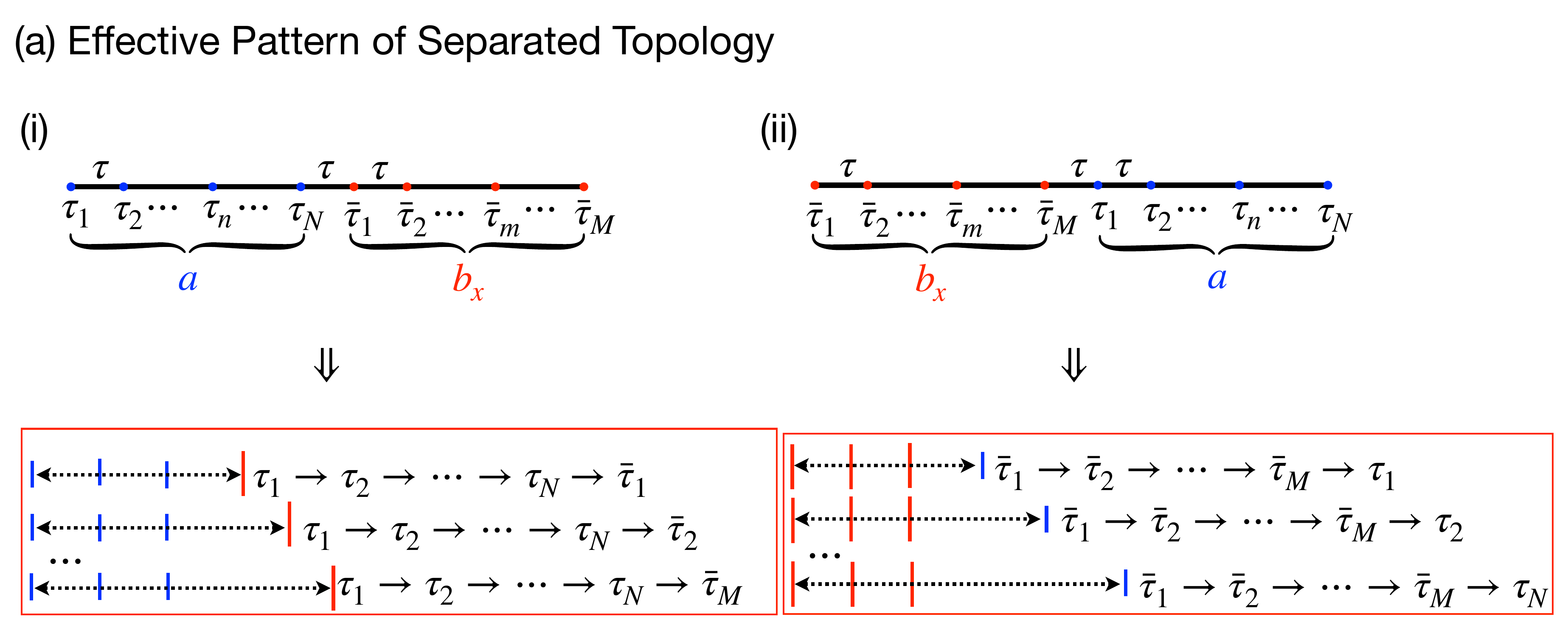}}
\subfigure{\includegraphics[width=0.49\textwidth]{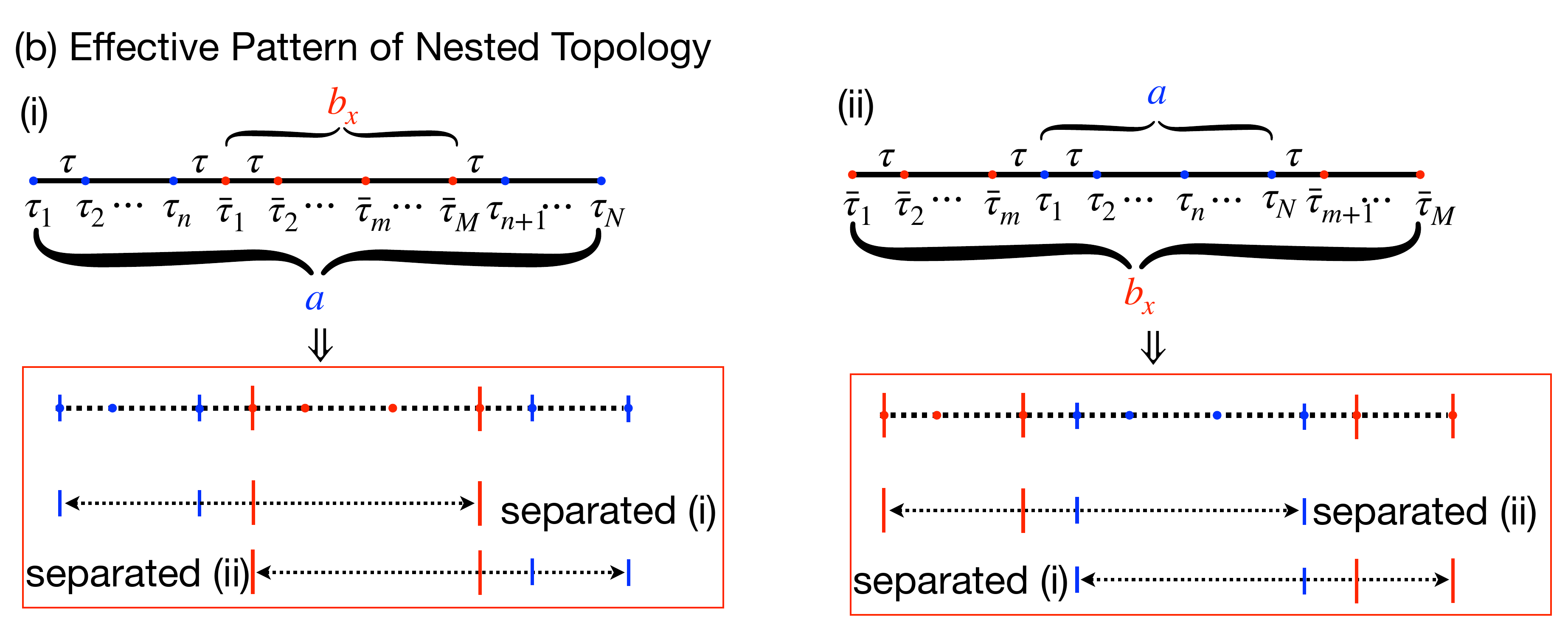}}
\subfigure{\includegraphics[width=0.49\textwidth]{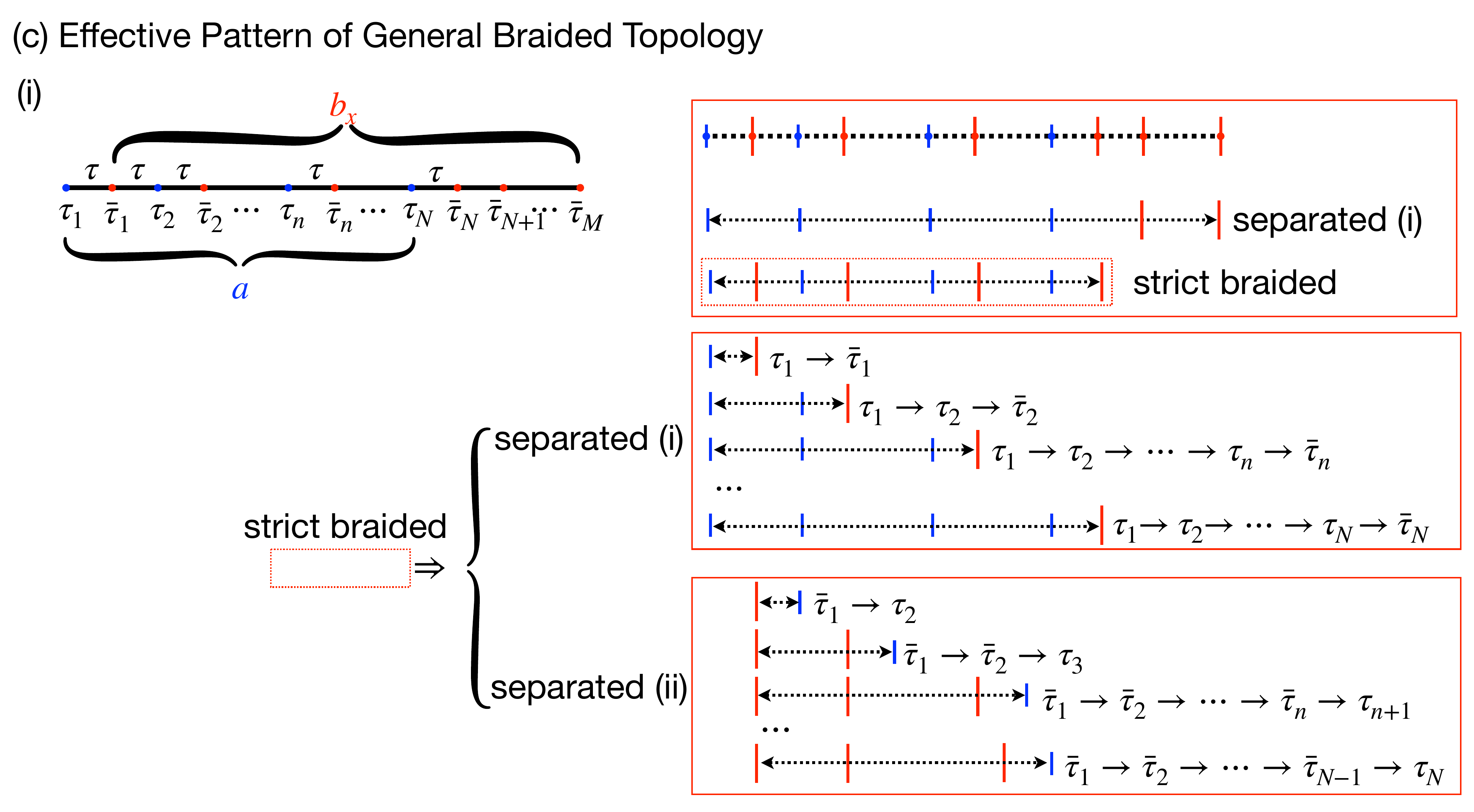}}
\subfigure{\includegraphics[width=0.49\textwidth]{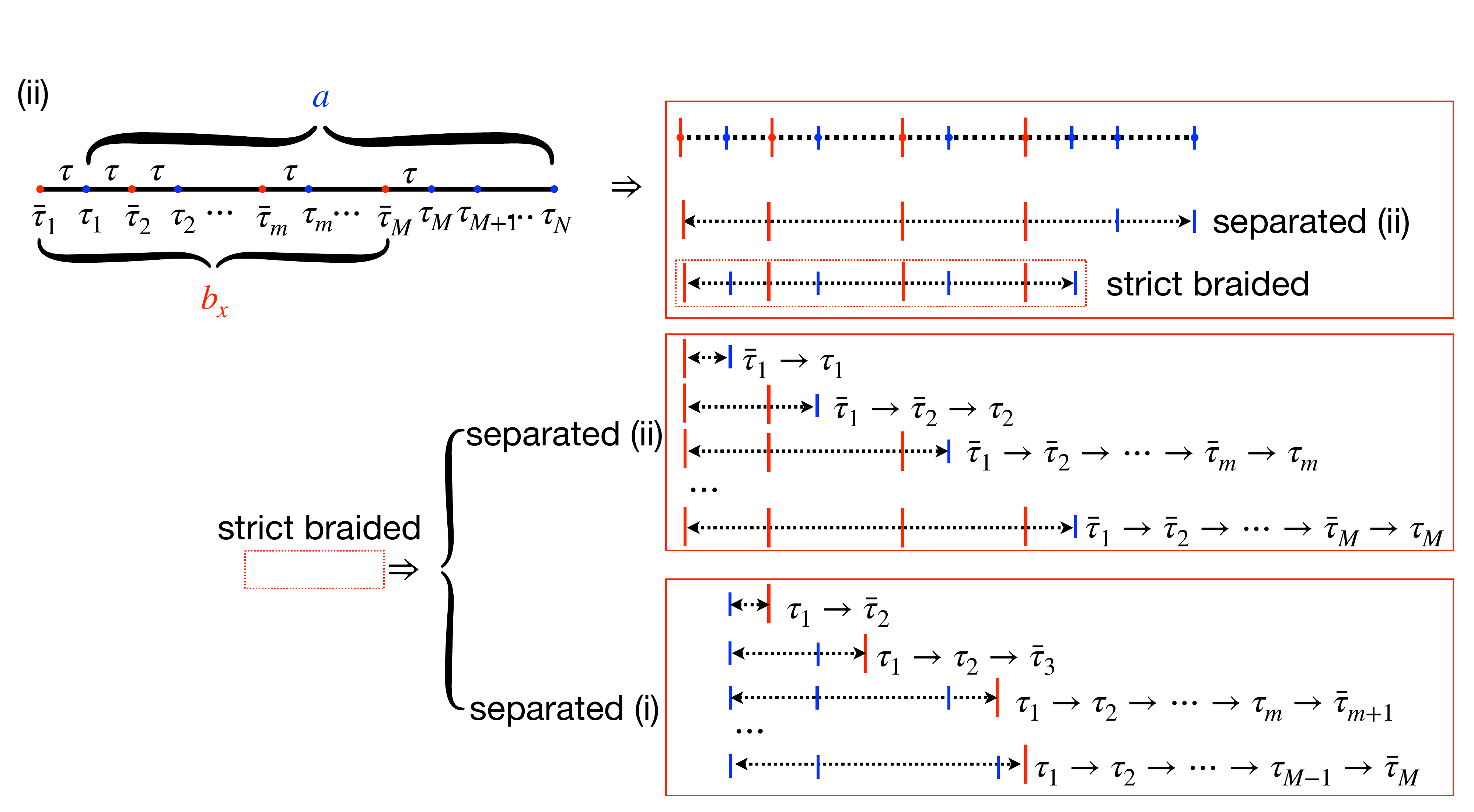}}
\caption{The effective patterns of topologies: (a) Separated, (b) Nested, and (c)  Braided. In (a), the separated topology is the most fundamental for the double-body system under consideration, while the other topologies can be viewed as combinations of different separated topologies. In (b), the nested topology can be categorized as a combination of separated topology (i) and separated topology (ii). In (c), the general braided topology can be divided into a strict braided component and a separated component. In the strict braided component, the number of interlaced coupling points is $N = M$, and this component can be further divided into two sets of separated topologies (i) and (ii).}
\label{EffectiveTop}
\end{figure*}
\subsection{Topologies in the considered double-body system}
In the previous section, we presented the transfer function $G(\omega)$. We now examine the elemetents $G_{21}(\omega)$ and $G_{12}(\omega)$, which represent a pair of cross responses: $\alpha_{\mathrm{in}}(\omega)\leadsto\beta_{\mathrm{out}}(\omega)$ and $\beta_{\mathrm{in}}(\omega)\leadsto\alpha_{\mathrm{out}}(\omega)$. According to Eq. \eqref{TransferFunction}, these two elements are proportional to the matrix $\mathcal{A}_g(\omega)$, and depend on the topologies that describe the relative positions of the modes ${a}$ and $b_x$. For the double-body system considered in this paper, the topologies can be categorized into three types: (a) separated, (b) nested, and (c) braided. {Although the topologies in multiple-point-coupling systems have been studied in previous works \cite{Kockum2018, Feng2021, Zhu2022A}, these studies focus only on two-point coupling cases, e.g., $N=M=2$, which do not consider topologies involving a greater number of coupling points. Moreover, most of these existing works \cite{Kockum2018, Feng2021} employ the Markovian approximation. As mentioned in Sec. \ref{Sec3}, this approach would manually break the inherent non-Markovian effects, thereby obscuring the underlying directional coupling and nonreciprocity. To address this limitation, we extend the analysis to include more coupling points and consider all possible topologies, and preserving the system's non-Markovian effects to explore topology-dependent nonreciprocity.} All possible topologies are illustrated in Fig. \ref{topology}, where the topologies labeled (ii) are mirror images of those labeled (i).

Here,  the topologies are subject to the following constraints:
 \begin{itemize}
 \item The coupling points associated with different modes do not overlap, which necessitates the existence of a time delay between modes $a$ and $b_x$. This is intended to ensure that the non-Markovian effects resulting from the time delay do not disappear in the limit Markovian regime, as discussed in the last paragraph of Sec. \ref{Sec2};
 \item The coupling points are distributed in the order $\tau_{n-1}<\tau_n<\tau_{n+1}, \bar{\tau}_{m-1}<\bar{\tau}_m<\bar{\tau}_{m+1}$;
\item The time delay is fixed as $\tau$ for the nearest-neighbor coupling points, and it is incorporated into the phase $\phi=(\omega_d-\omega)\tau$.
 \end{itemize}
Under the aforementioned constraints, although the coupling matrix $\mathcal{A}_g(\omega)$ differ acorss various topologies, they all represent the sum of a geometric sequence.  For simplicity, we present the detailed derivations in Appendix \ref{AppendixB}.

\noindent\textbf{(a) Separated.} The separated topology is the most fundamental, while the others can be viewed as various combinations of multiple separated topologies. Its effective pattern is shown in Fig. \ref{EffectiveTop} (a).

\noindent\textbf{(b) Nested.} Its effective pattern is shown in Fig. \ref{EffectiveTop} (b), which is categorized into two types: separated (i) and separated (ii). In the nested type (i), the coupling points $\bar{\tau}_{1} \sim \bar{\tau}_M$ are embedded between the points $\tau_n$ and $\tau_{n+1}$, necessitating that $N - n \geq 1$. In the mirroring topology (ii), the coupling points ${\tau}_{1} \sim {\tau}_N$ are embedded between the points $\bar{\tau}_m$ and $\bar{\tau}_{m+1}$, thereby requiring that $M - m \geq 1$.

\noindent\textbf{(c) Braided.} Strictly speaking, the braided topology illustrated in Fig. \ref{topology} should be referred to as the general braided topology, since some coupling points remain uninterlaced when $N\neq M$. In this work, we designate the braided topology as strict braided when $N=M$. The equivalent pattern of the general braided topology is depicted in Fig. \ref{EffectiveTop} (c), which can be divided into a strict braided component and a separated component. Moreover, the strict braided component is further categorized into two sets of separated topologies: (i) and (ii). To distinguish between strict and general braided topologies, we use the superscripts ``S" and ``G" to label the ``strict" and ``general" respectively. In the main text, we focus on the strict braided topology, while the general braided topology is discussed in Appendix \ref{AppendixC} as supplemental material.

\subsection{Nonreciprocities under different topologies}
Based on Eq. \eqref{CMatrices} and Eqs. \eqref{SeparatedAg}-\eqref{StrictBraidedAg}, it can be observed that the elements $\mathcal{A}_{21}(\omega)$ and $\mathcal{A}_{12}(\omega)$ are not always equal. This discrepancy results in unequal transfer function elements, specifically $|G_{21}(\omega)| \neq |G_{12}(\omega)|$, which indicates imbalanced responses between $\alpha_{\mathrm{in}}(\omega)\leadsto\beta_{\mathrm{out}}(\omega)$ and $\beta_{\mathrm{in}}(\omega)\leadsto\alpha_{\mathrm{out}}(\omega)$. This phenomenon is known as nonreciprocal transfer.

According to the \textit{reciprocity theorem}, a system is considered reciprocal if \textit{its transfer magnitude remains unchanged when the source and receiver are exchanged} \cite{Deak2012, Lau2018}. Consequently, nonreciprocity is the opposite concept. Reciprocity applies to multiple-input-multiple-output systems and requires the transfer matrix to satisfy the condition $|G_{ij}(\omega)|=|G_{ji}(\omega)|$. \footnote{This does not imply that the transfer function must be complex symmetric. A symmetric transfer function indicates that the dimensions of the output and input vectors are equal. More commonly, the dimension of the input vector may exceed that of the output vector when the system involves control loops, where additional inputs are utilized as control sources.} Furthermore, we emphasize that the transfer function describes how outputs respond to inputs. Therefore, the unequal matrix elements $|G_{12}(\omega)|$ and $|G_{21}(\omega)|$ indicate imbalanced transfers between the pathways $\alpha_{\mathrm{in}}\rightarrow a\rightarrow b_x\rightarrow\beta_{\mathrm{out}}$ and $\beta_{\mathrm{in}}\rightarrow b_x\rightarrow a\rightarrow \alpha_{\mathrm{out}}$. This imbalance can be evaluated by the nonreciprocal strength
\begin{equation}
\begin{aligned}
\sigma(\omega)&=\frac{|G_{21}(\omega)|^2-|G_{12}(\omega)|^2}{|G_{21}(\omega)|^2+|G_{12}(\omega)|^2}\\
&=\frac{|\mathcal{A}_{21}(\omega)|^2-|\mathcal{A}_{12}(\omega)|^2}{|\mathcal{A}_{21}(\omega)|^2+|\mathcal{A}_{12}(\omega)|^2},
\end{aligned}
\label{NonReciprocityStrength}
\end{equation}
where we have calculated the elements $G_{21}(\omega)$ and $G_{12}(\omega)$ in the second line via the Eq. \eqref{TransferFunction}. Clearly, the nonreciprocal strength $\sigma(\omega)\in[-1,1]$ describes three types of nonreciprocity:
\begin{itemize}
\item The \textit{right nonreciprocity}, denoted as $\sigma(\omega) \in (0, 1]$, indicates that the $\alpha_{\mathrm{in}}\leadsto \beta_{\mathrm{out}}$ response is greater than the $\beta_{\mathrm{in}}\leadsto \alpha_{\mathrm{out}}$ response. As the nonreciprocal strength $\sigma(\omega)$ varies from $0$ to $1$, the right nonreciprocity is enhanced, and the $\beta_{\mathrm{in}}\leadsto \alpha_{\mathrm{out}}$ response is completely truncated when $ \sigma(\omega) = 1$.
\item The \textit{left nonreciprocity}, denoted as $\sigma(\omega)\in[-1,0)$, which is the opposite of right nonreciprocity.
\item The \textit{reciprocity}, denoted as $\sigma(\omega)=0$, which means a balanced transfer between the two responses.
\end{itemize}

The nonreciprocity in quantum giant systems is a fascinating subject, yet it has received limited attention. To our knowledge, only one study \cite{Chen2022} involves this topic, examining nonreciprocity (also referred to as chirality) in a three-level giant atom. In their configuration, the three-level atom can be modeled as two coupled two-level atoms, each interacting with a waveguide at distinct locations. Notably, the nonreciprocity observed in their study arises from the preset phases $\theta_1$ and $\theta_2$ rather than from directional coupling. In contrast, the nonreciprocity in our system primarily results from structural asymmetry, as detailed in Eqs. \eqref{SeparatedAg}, \eqref{NestedAg} and \eqref{StrictBraidedAg}. This asymmetry facilitates directional coupling between the modes, which manifests as nonreciprocity.

Based on the coupling matrix $\mathcal{A}_g(\omega)$ (Eqs. \eqref{SeparatedAg}, \eqref{NestedAg} and \eqref{StrictBraidedAg}), one can readily determine the nonreciprocal strength $\sigma(\omega)$ of the topologies shown in Fig. \ref{topology}. {Besides the topologies, it also depends on two paparmeters: the phase $\phi$ and the number of coupling points $N, M$. A special case is the separated topology, for which the nonreciprocal strength is consistently $-1$ or $1$, regardless of these two parameters, c.f. Eq. \eqref{SeparatedAg}. For other topologies, the detailed nonreciprocal strength $\sigma(\omega)$ are decipted as Eqs. \eqref{NestedCondition} and \eqref{StrictCondtion}, with which one acquire nonreciprocity by tuning the phase $\phi$ for the given number of coupling points $N, M$.}

\begin{figure} [bt]
\centering
\includegraphics[width=0.46\textwidth]{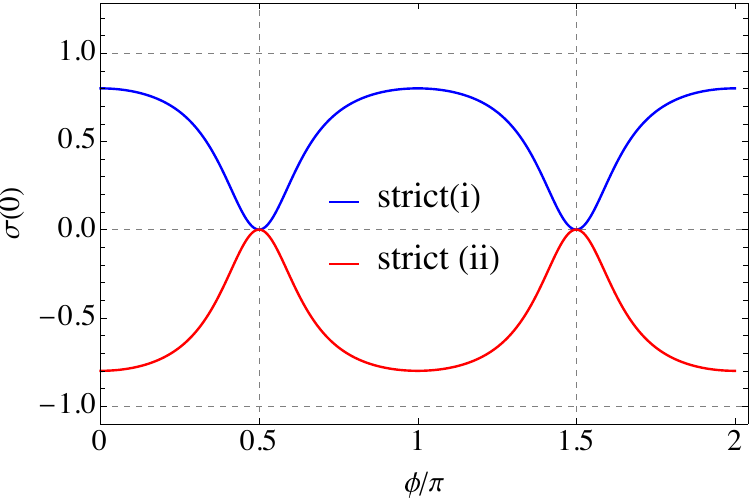}
\caption{(Color online) The numerical simulation of the nonreciprocal strength $\sigma(0)$ under strict braided topologies. The gridlines indicate the points $\sigma(0)=\{-1,0,1\}$. Due to destructive interference, the system exhibits reciprocity at the phases $\phi=\{0.5\pi, 1.5\pi\}$. }
\label{SigmaFig1}
\end{figure}

To encompass the topologies discussed in Ref. \cite{Kockum2018}, we set $N=M=2$. Besides, we also provide a detailed discussion on the number of coupling points in Appendix \ref{AppendixC}.  Now, let us return to the main topic and analyze the nonreciprocal strength $\sigma(\omega)$ of the topologies depicted in Ref. \cite{Kockum2018}. For brevity, we  assume that the system is resonantly driven to a steady state with $\Delta_1= \Delta_2=\omega =0$.  Under these assumptions, the nonreciprocal strength $\sigma(0)$ becomes
\begin{subequations}
\begin{eqnarray}
&&\sigma(0)=
\begin{cases}
1 & separated~(\mathrm{i}) \\
0 & nested~(\mathrm{i}) \\
\frac{2+2\cos 2\phi}{3+2\cos 2\phi} & strict~braided~(\mathrm{i})
\end{cases}\label{GSigmaA}\\
&&\sigma(0)=
\begin{cases}
-1 & separated~(\mathrm{ii}) \\
0 & nested~(\mathrm{ii}) \\
-\frac{2+2\cos 2\phi}{3+2\cos 2\phi} & strict~braided~(\mathrm{ii})
\end{cases}\label{GSigmaB}
\end{eqnarray}
\end{subequations}
In the separated topologies (i) and (ii), Eqs. \eqref{GSigmaA} and \eqref{GSigmaB} align with our intuition: energy is unidirectionally transferred from mode $a$ to  mode $b_x$ when mode $a$ is at the former position, and vice versa. Furthermore, the system consistently exhibits reciprocity in both nested topologies (i) and (ii). The underlying reason for this reciprocity is the balanced transfer between the pathways $a\rightarrow b_x$ and $b_x\rightarrow a$, rather than interference effects. For instance, in nested topology (i), the $a\rightarrow b_x$ transfer occurs through two pathways: $\tau_1 \rightarrow \bar{\tau}_1$ and $\tau_1 \rightarrow \bar{\tau}_2$, c.f. Fig. \ref{EffectiveTop} (b) . Conversely, the transfer $b_x\rightarrow a$ is facilitated by the pathways $\bar{\tau}_1 \rightarrow \tau_2$ and $\bar{\tau}_2 \rightarrow \tau_2$. The pathway $\tau_1 \rightarrow \bar{\tau}_1$ contributes equally to the pathway $\bar{\tau}_2 \rightarrow \tau_2 $, as do the pathways $\tau_1 \rightarrow \bar{\tau}_2 $ and $\bar{\tau}_1 \rightarrow \tau_2$. This symmetry results in a balanced energy transfer between the pathways $a \rightarrow b_x $ and $b_x \rightarrow a$, ensuring that the system remains reciprocal.

As illustrated in Fig. \ref{SigmaFig1}, the nonreciprocal strengths $\sigma(0)$ exhibit symmetry about the $\sigma(0) = 0$ axis, as the strict braided topologies (i) and (ii) are mutually mirroring. It is important to note that the points of interest occur at the phases $\phi = \{0.5\pi, 1.5\pi\}$, where the system maintains reciprocity. This phenomenon arises due to destructive interference among the various transfer pathways. For example, in strict braided topology (i), destructive interference occurs between the transfer pathways $ \tau_1 \rightarrow \bar{\tau}_1$ and $\tau_1 \rightarrow \bar{\tau}_2$. In strict braided topology (ii), destructive interference takes place between the pathways $\bar{\tau}_1 \rightarrow \tau_1$ and $ \bar{\tau}_1 \rightarrow \tau_2$. This destructive interference enforces a balanced energy transfer between the two modes. Aside from these two specific phases, the system is generally nonreciprocal; however, it does not achieve complete nonreciprocity of -1 or 1. This indicates that the transfer magnitude in one direction is consistently greater than that in the opposite direction, even though the energy transfer between the two modes remains reversible.

Before the end of this section, we quote the results in Appendix \ref{AppendixC} to briefly illustrate the impact of the number of coupling points on the nonreciprocal strength $\sigma(0)$. As the numbers of coupling points $N$ and $M$ increase, the separated component introduces additional phases $\mathcal{A}^A_{g}(0)$ to the matrix $\mathcal{A}_{g}(0)$, c.f. Eq. \eqref{Type3G}. These additional phases also contribute to the interferences, complicating the interference and altering the reciprocal points.

In the end, we provide a brief summary of the nonreciprocity induced by multiple-point couplings:

\begin{itemize}

\item {The nonreciprocity in multiple-point coupling systems is an intrinsic property arising fundamentally from directional couplings induced by time delays, as described in Eq. \eqref{Eom}. It represents a manifestation of non-Markovian effects. Consequently, the Markovian approximation is inapplicable to multiple-point coupling systems regardless of the time delay magnitude, since this approximation artificially eliminates the non-Markovian behavior. Therefore, as long as the coupling points do not overlap, nonreciprocity must persist. This characteristic implies that it is unnecessary to etch an excessively long waveguide to achieve pronounced non-Markovian effects and nonreciprocity, thereby facilitating on-chip integration of the system.}

\item Although directional coupling can induce nonreciprocal transfers, this does not necessarily imply that the system is always nonreciprocal. Under certain time delays, reciprocity can be restored due to destructive interference among various transfer pathways (see Figs. \ref{SigmaFig1} and \ref{SigmaFig2}). This destructive interference indicates the disappearance of the non-Markovian effect.

\item {The nonreciprocity depends on the topology, allowing one to adjust the structural parameters to achieve varying degrees of nonreciprocal strength. These parameters include the relative positions of the cavities, the number of coupling points, the distances between these points, and the coupling strengths to the waveguide.}
\end{itemize}
\begin{figure*} [bt]
\centering
\includegraphics[width=0.92\textwidth]{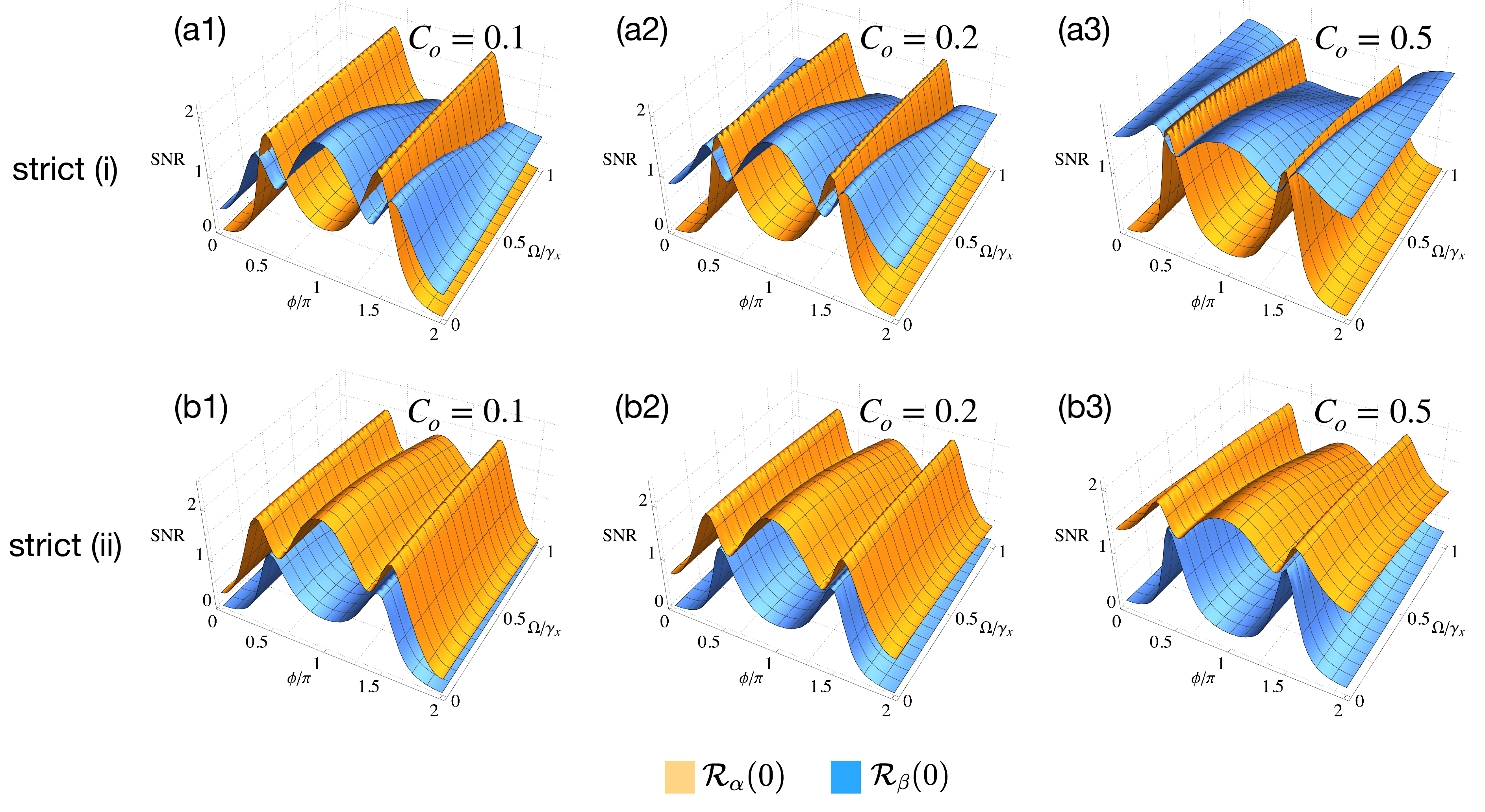}
\caption{(Color online) The numerical simulations of the SNRs $\mathcal{R}_{\alpha} (0)$ and $\mathcal{R}_{\beta} (0)$ in strict braided topologies. Parameters used for plotting are: $\gamma_x=\gamma_y$ and $\kappa=10\gamma_x$. Due to the residual term $\gamma_x+\frac{4\Omega^2}{\gamma_y}$, the SNRs $\mathcal{R}_{\alpha} (0)$ and $\mathcal{R}_{\beta} (0)$ do not strictly overlap at the reciprocal points $\phi=\{0.5\pi, 1.5\pi\}$. In panels (a1)-(a3), the system exhibits right nonreciprocity, except at the reciprocal points, resulting in the SNR $\mathcal{R}_{\beta} (0)$ being greater than the SNR $\mathcal{R}_{\alpha} (0)$. More importantly, the SNR $\mathcal{R}_\beta (0)$ in the interval $\phi \in [0.5\pi, 1.5\pi]$ varies synchronously with the nonreciprocal strength $\sigma(0)$, as shown in Fig. \ref{SigmaFig1} (a). This synchronous variation reveals that nonreciprocity can effectively enhance the SNR. Similar results are also observed in the mutual mirroring topology (b1)-(b3), where the roles of the SNRs $\mathcal{R}_\beta (0)$ and $\mathcal{R}_\alpha (0)$ are reversed.}
\label{SNR1}
\end{figure*}

\section{Impacts of nonreciprocity on sensing performances\label{Sec5}}
\subsection{The impact of nonreciprocity on SNR}
After discussing nonreciprocity, we will now focus on the role it plays in sensory perception. This process relies on the output signal
\begin{eqnarray}
\mathcal{S}_{\epsilon}(\omega)=|\langle\epsilon_{\mathrm{out}}(\omega)\rangle|^2,
\label{Signal}
\end{eqnarray}
where $\epsilon=\{\alpha, \beta\}$ and the symbol $\langle \cdot\rangle$ means taking the expectation value. In experiments, coherent lasers are typically utilized as probe fields. Consequently, the probe field $\epsilon_{\mathrm{in}}$ can be expressed as $\epsilon_{\mathrm{in}}=\epsilon+\epsilon^q_{\mathrm{in}}(t)$, where $\epsilon$ represents the classical amplitude and $\epsilon^q_{\mathrm{in}}$ signifies the quantum vacuum input \cite{Scully1997, Clerk2010}. {Additionally, the system often operates within a dilution refrigerator at temperatures approaching absolute zero \cite{Noguchi2020, Wu2021}. This further leads to the following correlations of the quantum input vector}
\begin{subequations}
\begin{eqnarray}
&&\langle \delta \tilde{\boldsymbol{\mathrm{u}}}_{\mathrm{in}}(\omega)\rangle = \langle \delta \tilde{\boldsymbol{\mathrm{u}}}^\dag_{\mathrm{in}}(\omega)\rangle=0, \label{Correlation1}\\
&&\langle \delta \tilde{\boldsymbol{\mathrm{u}}}_{\mathrm{in}}^\dag(\omega)\delta \tilde{\boldsymbol{\mathrm{u}}}_{\mathrm{in}}(\omega')\rangle = 0,\label{Correlation2}\\
&&\langle \delta \tilde{\boldsymbol{\mathrm{u}}}_{\mathrm{in}}(\omega)\delta \tilde{\boldsymbol{\mathrm{u}}}_{\mathrm{in}}^\dag(\omega')\rangle = 2\pi\delta(\omega-\omega')I_{5\times5},\label{Correlation3}
\end{eqnarray}
\end{subequations}
where $\delta O=O-\langle O\rangle$ denotes the fluctuation of an arbitrary operator $O$. Besides, the output signal is surrounded by the quantum noise
\begin{eqnarray}
\mathcal{N}_{\epsilon}(\omega)&=&\frac{1}{2}\int\mathrm{d}t~e^{-i\omega t}\langle [\delta \varepsilon_{\mathrm{out}}(t),\delta \varepsilon^\dag_{\mathrm{out}}(0)]_+\rangle\nonumber\\
&=&\frac{1}{2}\frac{1}{2\pi}\int\mathrm{d}\omega' \langle [\delta\varepsilon_{\mathrm{out}}(\omega), \delta\varepsilon^\dag_{\mathrm{out}}(\omega')]_+\rangle,\label{Noise}
\end{eqnarray}
where the anti-commutation relation reads $[O, O^\dag]_+=OO^\dag+O^\dag O$. Based on Eq. \eqref{EnergyConservation}, the quantum noise satisfies
\begin{eqnarray}
\mathcal{N}_{\epsilon}(\omega)\equiv\frac{1}{2}.\label{ShotNoiseLimit}
\end{eqnarray}
The constant $\frac{1}{2}$ represents the so-called shot noise, which delineates the minimum limit of quantum noise permitted by the Heisenberg uncertainty relation \cite{Lau2018, Clerk2010}. {This limit can be attained only when the quantum input vector is in the vacuum state; the zero-Kelvin state is also considered as the vacuum state. }Unlike optomechanical systems \cite{Zhu2022B}, the shot noise limit in this context cannot be surpassed by configuring the input $ \varepsilon^q_{\mathrm{in}}(\omega)$ to a squeezed state $|\xi\rangle_{\varepsilon}=e^{-\frac{1}{2}(\xi \varepsilon^{q\dag}_{\mathrm{in}}(\omega)\varepsilon^{q\dag}_{\mathrm{in}}(\omega)-\xi^*\varepsilon^{q}_{\mathrm{in}}(\omega)\varepsilon^q_{\mathrm{in}}(\omega))}|0\rangle$. However, this does not imply that our system exhibits worse noise compared to optomechanical systems. In optomechanical systems \cite{Zhu2022B}, the minimum noise after squeezing, given by $\mathcal{N}_{\varepsilon}(\omega)= \frac{1}{2}[e^{-2|\xi|}+(e^{|\xi|}+1)^2-1]\geq\frac{3}{2}$, remains greater than the shot noise limit \eqref{ShotNoiseLimit}.

Using Eqs. \eqref{Signal} and \eqref{Noise}, one can further derive the SNR per photon, which is one of the most important indices of sensors
\begin{subequations}
\begin{eqnarray}
\mathcal{R}_{\alpha}(\omega)=\frac{\mathcal{S}_{\alpha}(\omega)}{\mathcal{N}_{\alpha}(\omega)|\alpha|^2}=2|G_{11}(\omega)+r e^{i\theta} G_{12}(\omega)|^2
\label{SNRAlpha}\\
\mathcal{R}_{\beta}(\omega)=\frac{\mathcal{S}_{\beta}(\omega)}{\mathcal{N}_{\beta}(\omega)|\alpha|^2}=2|G_{21}(\omega)+r e^{i\theta}G_{22}(\omega)|^2
\label{SNRBeta}
\end{eqnarray}
\end{subequations}
with the amplitude ratio $r=|\beta/\alpha|$ and the phase $\theta=\mathrm{arg} (\beta/\alpha)$. The advantage of the definition on SNR lies in the fact that the performance of the gyroscope depends solely on the system itself, rather than on external probes. Notably, the SNR per photon provides a readable condition $\mathcal{R}_{\epsilon}(\omega)\geq1$; otherwise the output signal is overwhelmed by noise \cite{Zhu2024}. Unlike single-input-single-output systems \cite{Zhu2024}, this condition is easier to achieve in double-input-double-output system, as one can consistently obtain a higher SNR per photon by adjusting the ratio $r=|\beta/\alpha|$ without altering the total input photon number $n_{\mathrm{in}}=|\alpha|^2+|\beta|^2$. For fair comparisons, we set $r=1, \theta=0, \kappa_a=\kappa_b=\kappa$ in the following sections.

 \begin{figure*} [bt]
\centering
\includegraphics[width=0.92\textwidth]{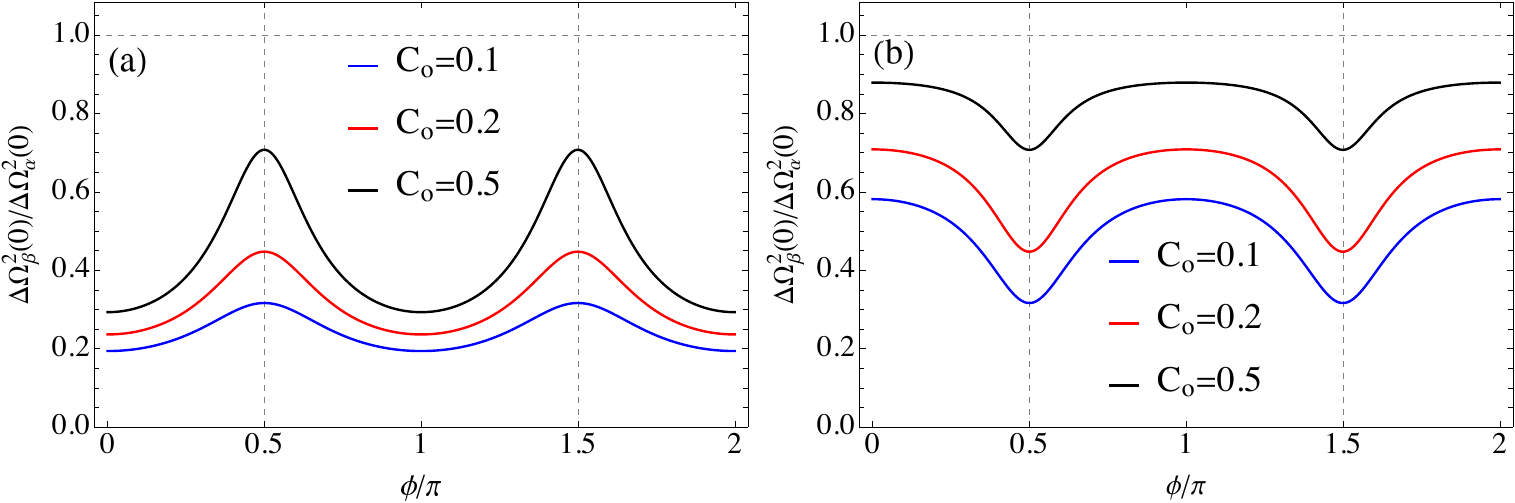}
\caption{(Color online) The numerical simulations of the sensitivity ratio $\Delta\Omega_\beta^2(0)/\Delta\Omega_\alpha^2(0)$ under (a) strict braided topology (i) and (b) strict braided topology (ii). In (a), due to the advantageous effects of right nonreciprocity, the sensitivity ratio $\Delta\Omega_\beta^2(0)/\Delta\Omega_\alpha^2(0)$ is less than 1 and reaches its minimum at the maximum nonreciprocal points $\phi=\{0,\pi,2\pi\}$. In (b), the sensitivity ratio $\Delta\Omega_\beta^2(0)/\Delta\Omega_\alpha^2(0)$ remains less than 1, indicating that the output signal $\mathcal{S}_\alpha(0)$ does not benefit from left nonreciprocity in the sensitivity $\Delta\Omega_\alpha^2(0)$. This limitation arises because the system operates within the weak coupling regime.}
\label{Sensitivity1}
\end{figure*}

As a gyroscope, its purpose is to readout the unknown angular velocity $\Omega$ through its output signals. Consequently, these output signals must inherently reflect the angular velocity, regardless of the system's parameters. This requirement narrows the feasible topologies to only the strict and general braided configurations that satisfy this condition. In contrast, the output signals in other topologies become independent of the angular velocity under certain time delays. Under the same assumptions $\omega=\Delta_1=\Delta_2=0$, the SNRs per photon in the strict braided topologies read
\begin{subequations}
\begin{eqnarray}
&&\mathcal{R}_\alpha(0)=
\begin{cases}
2|1-\frac{f^s_1\kappa+\gamma_x+\frac{4\Omega^2}{\gamma_y}}{F^s_1\frac{\kappa}{2}+F^s_2(\frac{\gamma_x}{2}+\frac{2\Omega^2}{\gamma_y})]}|^2 & strict~\mathrm{(i)}\\
2|1-\frac{f^s_2\kappa+\gamma_x+\frac{4\Omega^2}{\gamma_y}}{F^s_1\frac{\kappa}{2}+F^s_2(\frac{\gamma_x}{2}+\frac{2\Omega^2}{\gamma_y})]}|^2 & strict~\mathrm{(ii)}\\
\end{cases}\label{RAStrict}\\
&&\mathcal{R}_\beta(0)=
\begin{cases}
2|1-\frac{f^s_2\kappa}{F^s_1\frac{\kappa}{2}+F^s_2(\frac{\gamma_x}{2}+\frac{2\Omega^2}{\gamma_y})]}|^2 & strict~\mathrm{(i)}\\
2|1-\frac{f^s_1\kappa}{F^s_1\frac{\kappa}{2}+F^s_2(\frac{\gamma_x}{2}+\frac{2\Omega^2}{\gamma_y})]}|^2 & strict~\mathrm{(ii)}\\
\end{cases}\label{RBStrict}
\end{eqnarray}
\end{subequations}
where $f^s_1=1+\sqrt{C_o}(1-e^{i\phi}+e^{2i\phi})$, $f^s_2=1+\sqrt{C_o}(1-2e^{i\phi}+e^{2i\phi}-e^{3i\phi})$, $F^s_1=1+C_o+2\sqrt{C_o}(1+e^{2i\phi})$, $F^s_2=1+\sqrt{C_o}(1+e^{2i\phi})$, and $C_o=4\gamma^2/\kappa^2$ is the double-photon cooperativity. According to experiments \cite{Zhu2024, Chu2018}, we consider the conditions $\Omega\sim\gamma_{x, y}\ll\kappa$ and $C_o\ll1$. These parameters correspond to a realistic scenario, where the gyroscope is implemented using a high-Q cavity and operates in the weak coupling regime, with the unknown angular velocity $\Omega$ treated as a weak perturbation. Under these conditions, the numerical simulations of the SNRs $\mathcal{R}_\alpha(0)$ and $\mathcal{R}_\beta(0)$ are provided in Fig. \ref{SNR1}.

When comparing Eq. \eqref{RAStrict} with Eq. \eqref{RBStrict}, a residual term $\gamma_x + \frac{4\Omega^2}{\gamma_y}$ appears in the numerator of $\mathcal{R}_\alpha(0)$. This causes the SNR $\mathcal{R}_\alpha(0)$ not to strictly overlap with $\mathcal{R}_\beta(0)$ at the reciprocal points $\phi = \{0.5\pi, 1.5\pi\}$, as shown in Fig. \ref{SNR1}. If this residual term is ignored, both SNRs overlap at the reciprocal points. In the strict braided topology (i), the system exhibits right nonrciprocity $\sigma(0)>0$ except at the reciprocal points $\phi = \{0.5\pi, 1.5\pi\}$, c.f. Fig. \ref{SigmaFig1}. This right nonreciprocity results in the SNR $\mathcal{R}_\beta(0)$ being globally greater than the SNR $\mathcal{R}_\alpha(0)$, as shown in Figs. \ref{SNR1} (a1)-(a3). Specifically, the SNR $\mathcal{R}_\beta(0)$ varies synchronously with the nonreciprocal strength $\sigma(0)$ within the interval $\phi \in [0.5\pi, 1.5\pi]$, and both increase to their maximum values at $\phi = \pi$. Compared to the reciprocal points $\phi=\{0.5\pi, 1.5\pi\}$, the right nonreciprocity enhances the SNR $\mathcal{R}_\beta(0)$ at the phase $\phi = \pi$. However, this enhancement does not imply that the right nonreciprocity amplifies the signal $\mathcal{S}_{\beta}(\omega)$, as the system conserves energy, c.f. Eq. \eqref{EnergyConservation}. Indeed, it functions as a valve, directing energy transfer more efficiently toward the $a \rightarrow b_x$ direction. Consequently, the signal $\mathcal{S}_\beta(0)$ is enhanced while the signal $\mathcal{S}_\alpha(0)$ is suppressed. This finding opens new possibilities for further improving the SNR, specifically by constructing a system with higher nonreciprocity to maximize the transfer of input to the desired port. Similar results also emerge in the mirroring topology (ii), but the roles of $\mathcal{R}_\beta(0)$ and $\mathcal{R}_\alpha(0)$ are reversed, as depicted in Figs. \ref{SNR1} (b1)-(b3).

\subsection{The impact of nonreciprocity on Sensitvity}
We now shift our attention to sensitivity, another important index for sensors. Sensitivity, often referred to as accuracy, denotes the minimum detectable change in the quantity being measured \cite{Wu2020,Guo2021,Koppenhofer2023, Zhu2024}. The sensitivity is defined as
\begin{eqnarray}
\Delta\Omega^2_\epsilon(\omega)=\lim_{\Omega^2\rightarrow0}\sqrt{|\frac{\mathcal{N}_\epsilon(\omega)}{\partial_{\Omega^2}\langle\epsilon_{\mathrm{out}}(\omega)\rangle}|^2}\label{Sensitivity}.
\end{eqnarray}
Notably, the sensitivity $\Delta\Omega^2_\epsilon(\omega)$ is expressed in units of $\mathrm{Hz}^2$, as the angular velocity $\Omega$ consistently appears in a squared form within the output signals $\mathcal{S}_{\varepsilon}(\omega)$ (c.f. Eqs. \eqref{RAStrict}-\eqref{RBStrict}). Under the conditions $\Omega\sim\gamma_{x, y}\ll\kappa$ and $\omega=\Delta_1=\Delta_2=0$ that we used previously, the sensitivity read
\begin{subequations}
\begin{eqnarray}
&&\Delta\Omega_\alpha^2(0)\simeq
\begin{cases}
\frac{\gamma_y\kappa}{16|\alpha|}|\frac{(F_1^s)^2}{f^s_2\sqrt{C_o}}| & strict~\mathrm{(i)}\\
\frac{\gamma_y\kappa}{16|\alpha|}|\frac{(F_1^s)^2}{f^s_1\sqrt{C_o}(2+e^{2i\phi})}| & strict~\mathrm{(ii)}
\end{cases}\label{SensitivityA}\\
&&\Delta\Omega_\beta^2(0)\simeq
\begin{cases}
\frac{\gamma_y\kappa}{16|\alpha|}|\frac{(F_1^s)^2}{f_2^sF_2^s}| & strict~\mathrm{(i)}\\
\frac{\gamma_y\kappa}{16|\alpha|}|\frac{(F_1^s)^2}{f_1^sF_2^s}| & strict~\mathrm{(ii)}.
\end{cases}\label{SensitivityB}
\end{eqnarray}
\end{subequations}
To illustrate the effects of nonreciprocity, we plot the sensitivity ratio $\Delta\Omega_\beta^2(0)/\Delta\Omega_\alpha^2(0)$ in Fig. \ref{Sensitivity1}. In the figure, labels (a) and (b) denote the strict braided topologies (i) and (ii), respectively.

\begin{figure*} [bt]
\centering
\includegraphics[width=0.92\textwidth]{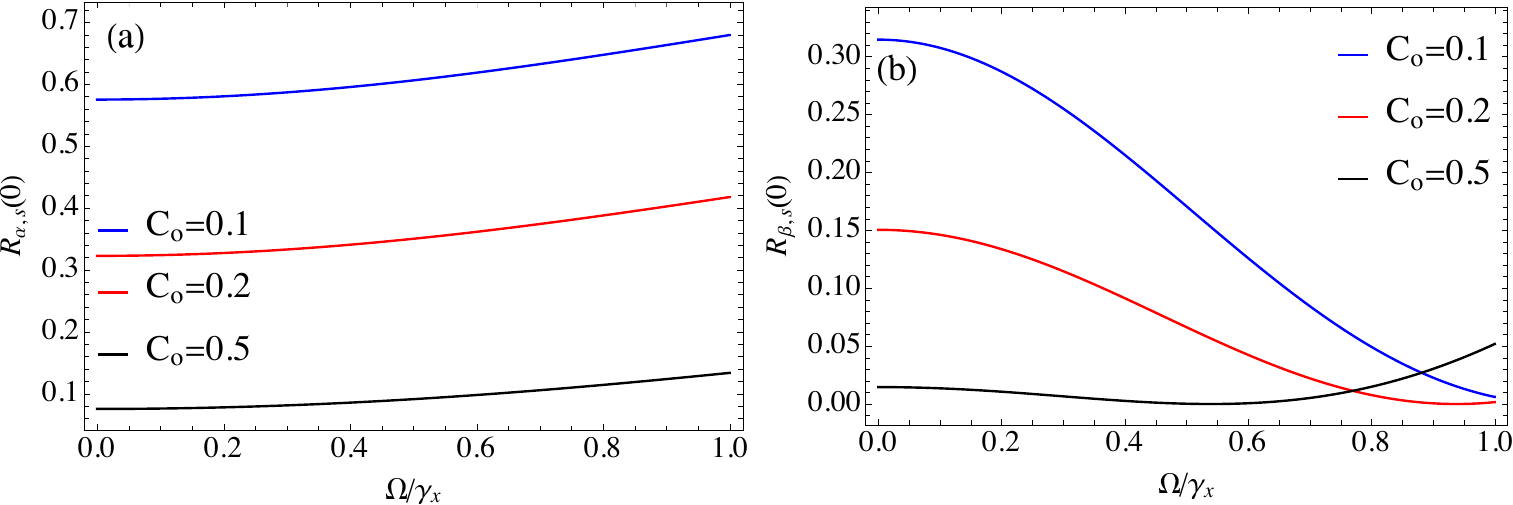}
\caption{(Color online) The numerical simuations of the SNRs in the tranditional  system (i). Both the SNRs $\mathcal{R}_{\alpha, s}(0)$ and $\mathcal{R}_{\beta, s}(0)$ are less than 1, indicating that both output signals $\mathcal{S}_{\alpha, s}(0)$ and $\mathcal{S}_{\beta, s}(0)$ are unreadable.}
\label{ComparisonSNR}
\end{figure*}
\begin{figure*} [bt]
\centering
\includegraphics[width=0.92\textwidth]{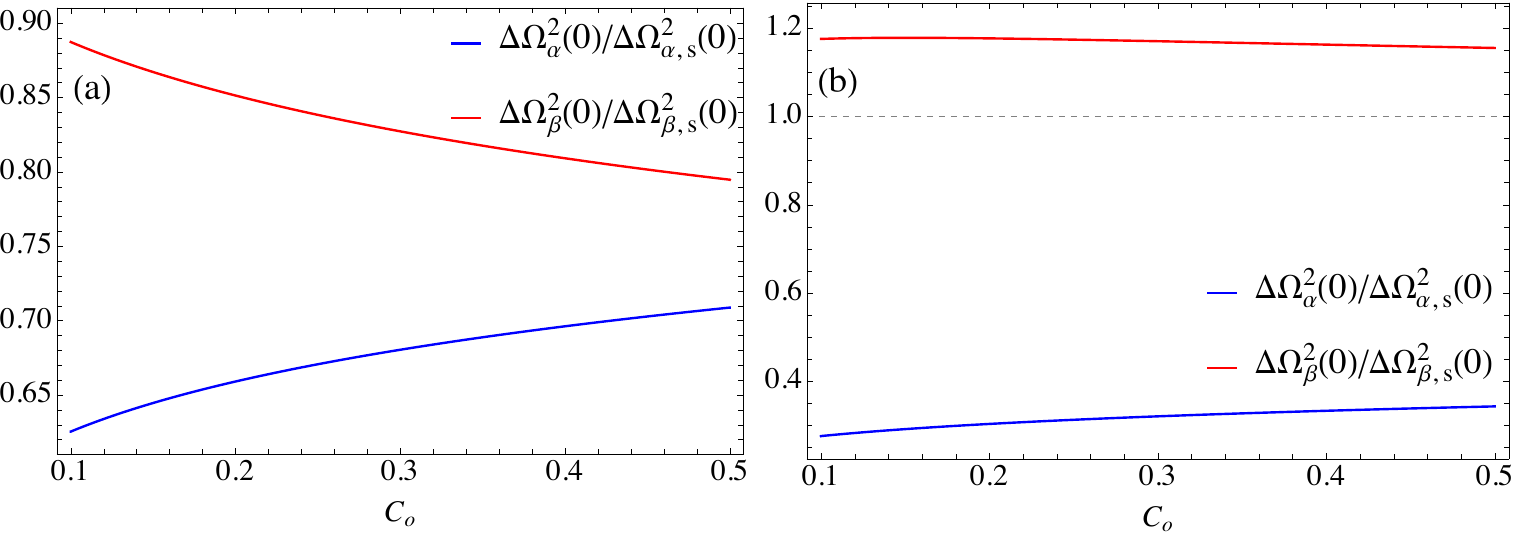}
\caption{(Color online) The comparison of the sensitivities between the multiple-point-coupling system and the traditional system (i). In (a), the sensitivity $\Delta\Omega_{\varepsilon}^2(0)$ is set at the phase $\phi=\pi$ and selected from the strict braided topology (i). The right nonreciprocity in the multiple-point-coupling structure induces the sensitivity $\Delta\Omega_{\varepsilon}^2(0)$ lower than sensitivity $\Delta\Omega_{\varepsilon, s}^2(0)$. In (b), the sensitivity $\Delta\Omega_{\varepsilon}^2(0)$ is also set at the phase $\phi=\pi$ but selected from the strict braided topology (ii). The improvements on the sensitivity led by the left nonreciprocity is reflected on the sensitivity $\Delta\Omega_{\alpha}^2(0)$.  }
\label{ComparisonSensitivity}
\end{figure*}

Benefiting from the right nonreciprocity, the sensitivity ratio $\Delta\Omega_\beta^2(0)/\Delta\Omega_\alpha^2(0)$ in the strict braided topology (i) consistently remains less than 1, regardless of the cooperativity. Especially it reaches minimum at the maximum nonreciprocal points $\phi=\{0, \pi, 2\pi\}$, as shown in Fig. \ref{Sensitivity1} (a). This reveals that the right nonreciprocity makes the output signal $\mathcal{S}_{\beta}(0)$ more sensitive to changes in angular velocity $\Omega$. In particular, this effect becomes more pronounced within the interval $\phi=[0.5\pi, 1.5\pi]$ as the cooperativity $C_o$ increases: the sensitivity ratio $\Delta\Omega_\beta^2(0)/\Delta\Omega_\alpha^2(0)$ decreases with the increasing nonreciprocal strength $\sigma(0)$. However, in the strict braided topology (ii), the sensitivity ratio $\Delta\Omega_\beta^2(0)/\Delta\Omega_\alpha^2(0)$ exhibits different trends, as shown in Fig. \ref{Sensitivity1} (b). Contrary to our expectations, it does not exceed 1 and reaches a minimum at the reciprocal points $\phi=\{0.5\pi, 1.5\pi\}$, indicating that the output signal $\mathcal{S}_\alpha(0)$ does not benefit from left nonreciprocity. The underlying reason is that a sensitivity ratio $\Delta\Omega_\beta^2(0)/\Delta\Omega_\alpha^2(0)>1$ requires the cooperativity $C_o>1$. If the cooperativity $C_o>1$ is satisfied, the output signal $\mathcal{S}_\alpha(0)$ would benefit from left nonreciprocity; that is, the sensitivity ratio $\Delta\Omega_\beta^2(0)/\Delta\Omega_\alpha^2(0)>1$ not only exceeds 1 but also exhibits a completely opposite trend compared to Fig. \ref{Sensitivity1} (a). However, the condition $C_o > 1$ lies beyond the weak coupling regime considered in this work. Therefore, strictly speaking, the results in Fig. \ref{Sensitivity1} (b) do not conflict with those in Fig. \ref{Sensitivity1} (a).

Based on the results presented in Figs. \ref{SNR1} and \ref{Sensitivity1}, we can conclude the significant role of nonreciprocity in sensing. It compels energy to transfer more concentratedly in one direction. In this direction, the output signal is enhanced, making it more sensitive to changes in the quantities being measured.

\section{Comparison with traditional proposals \label{Sec6}}

In Sec. \ref{Sec5}, we thoroughly discussed the role of nonreciprocity in sensing applications. Now, we will compare our gyroscope with traditional single-point-coupling gyroscopes to further illustrate the enhancements in sensing performance due to nonreciprocity. The traditional gyroscopes are realized by the following two structures: (i) both modes $a$ and $b_x$ couple to the waveguide at the same point, i.e., $N=M=1$ and $\tau=0$. (ii) mode $a$ directly couples to the mode $b_x$ with a Hermitian Hamiltonian $H_g \rightarrow H_d = i\frac{\sqrt{C_o}\kappa}{2}(a^\dag b_x - H.c.)$. In both cases, the system is reciprocal.

\noindent\textbf{Traditional System (i)}: The two modes couple to the waveguide at the same point. Under the same conditions $\Omega\sim\gamma_{x, y}\ll\kappa$ and $\omega=\Delta_1=\Delta_2=0$, the SNRs in this structure read
\begin{eqnarray}
&&\mathcal{R}_{\alpha, s}(0)=2|1-\frac{\kappa+\gamma_x+\frac{4\Omega^2}{\gamma_y}}{\frac{\kappa}{2}(1+\sqrt{C_o})+(1+\frac{\sqrt{C_o}}{2})(\frac{\gamma_x}{2}+\frac{2\Omega^2}{\gamma_y})}|^2\label{SNRsmallA}\nonumber\\
\\
&&\mathcal{R}_{\beta, s}(0)=2|1-\frac{\kappa}{\frac{\kappa}{2}(1+\sqrt{C_o})+(1+\frac{\sqrt{C_o}}{2})(\frac{\gamma_x}{2}+\frac{2\Omega^2}{\gamma_y})}|^2\label{SNRsmallB}\nonumber\\
\end{eqnarray}
which are simulated in Fig. \ref{ComparisonSNR}, where we use the subscript $s$ to denote the ``sing-point-coupling structure". Due to the residual term $\gamma_x+\frac{4\Omega^2}{\gamma_y}$, the SNRs $\mathcal{R}_{\alpha, s}(0)$ and $\mathcal{R}_{\beta, s}(0)$ do not exhibit the same values, even though the system is reciprocal. This result is analogous to Eqs. \eqref{RAStrict} and \eqref{RBStrict}. By comparing Fig. \ref{ComparisonSNR} with Fig. \ref{SNR1}, we observe that both SNRs $\mathcal{R}_{\alpha, s}(0)$ and $\mathcal{R}_{\beta, s}(0)$ are less than 1. This indicates that the output signals $\mathcal{S}_{\alpha, s}(0)$ and $\mathcal{S}_{\beta, s}(0)$ are unreadable under the same parameters. This finding underscores significant advancements in nonreciprocity, suggesting that it can transform unreadable signals into readable ones.

Correspondingly, the sensitivities take the form
\begin{eqnarray}
&&\Delta\Omega^2_{\alpha, s}(0)\simeq\frac{\gamma_y\kappa}{16|\alpha|}|\frac{2(1+\sqrt{C_o})^2}{\sqrt{C_o}}|\\
&&\Delta\Omega^2_{\beta, s}(0)\simeq\frac{\gamma_y\kappa}{16|\alpha|}|\frac{(1+\sqrt{C_o})^2}{1+\frac{\sqrt{C_o}}{2}}|.
\end{eqnarray}
Here, we utilize the ratio $\Delta\Omega^2_{\varepsilon}(0)/\Delta\Omega^2_{\varepsilon, s}(0)$ to demonstrate the enhancements in the sensitivities of nonreciprocity, as simulated in Fig. \ref{ComparisonSensitivity}. The sensitivity $\Delta\Omega^2_{\varepsilon}(0)$ in Fig. \ref{ComparisonSensitivity} (a) and (b) is set at the phase $\phi=\pi$ and is selected from the strict braided topologies (i) and (ii), respectively. Benefiting from right nonreciprocity, the sensitivity $\Delta\Omega^2_{\varepsilon}(0)$ in the strict braided topology (i) is lower than the sensitivity $\Delta\Omega^2_{\varepsilon, s}(0)$, as shown in Fig. \ref{ComparisonSensitivity} (a). In contrast to the single-point coupling structure, the nonreciprocity of the multiple-point coupling structure concurrently enhances the sensitivity of the output signals $\mathcal{S}_{\varepsilon}(0)$ in the strict braided topology (i). In particular, the maximum enhancement is approximately $38\%$ (corresponding to $\Delta\Omega^2_{\beta}(0)/\Delta\Omega^2_{\beta, s}(0)\approx 0.62$ at $C_o=0.1$). In the strict braided topology (ii), this enhancement is only reflected in the output signal $\mathcal{S}_{\alpha}(0)$, as shown in Fig. \ref{ComparisonSensitivity} (b). In this case, the sensitivity is maximally enhanced by $72\%$ (corresponds to $\Delta\Omega^2_{\alpha}(0)/\Delta\Omega^2_{\alpha, s}(0)\approx 0.28$ at $C_o=0.1$). Based on the results presented in Figs. \ref{ComparisonSNR} and \ref{ComparisonSensitivity}, we find that, compared to the reciprocal single-point coupling structure, the nonreciprocity introduced by the giant-cavity structure has a significant positive effect on sensing performance. It not only enhances the SNR, making previously unreadable signals readable, but also improves sensitivity.

\begin{figure*} [bt]
\centering
\includegraphics[width=0.92\textwidth]{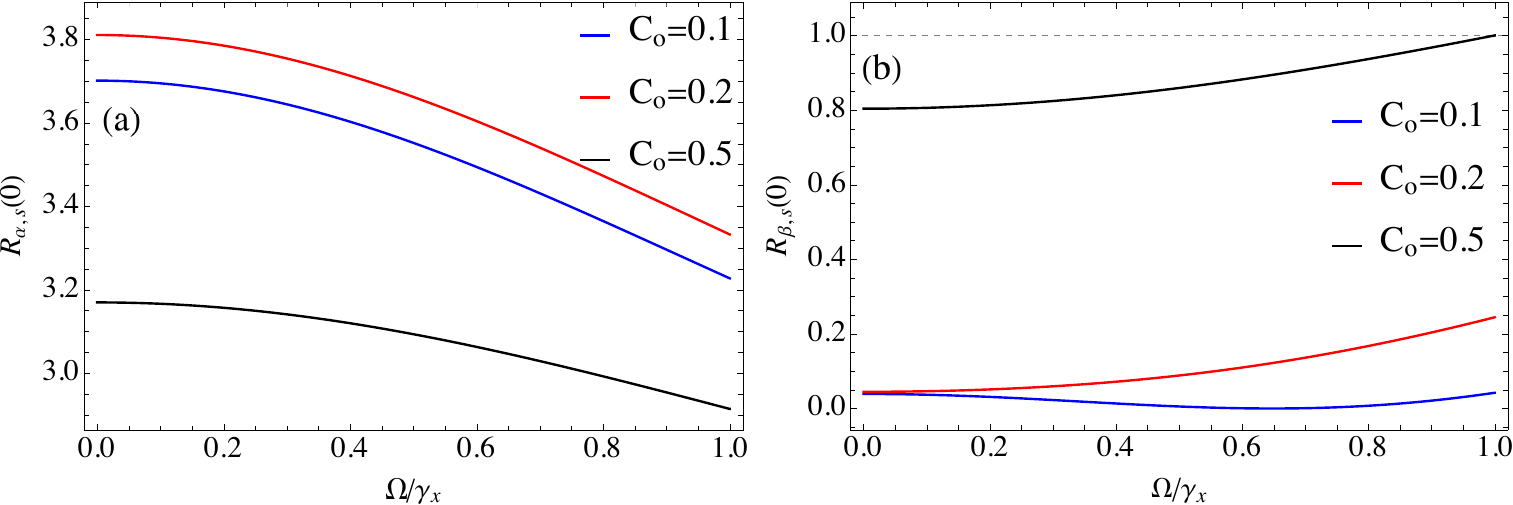}
\caption{(Color online) The numerical simuations of the SNRs in the tranditional  system (ii). Compared to Fig. \ref{SNR1}, only the output signal $\mathcal{S}_{\alpha, s} (0)$ in the coupling structure can be used for readout as only the SNR $\mathcal{R}_{\alpha, s} (0)$ is greater than 1.}
\label{ComparisonSNR2}
\end{figure*}
\begin{figure*} [bt]
\centering
\includegraphics[width=0.92\textwidth]{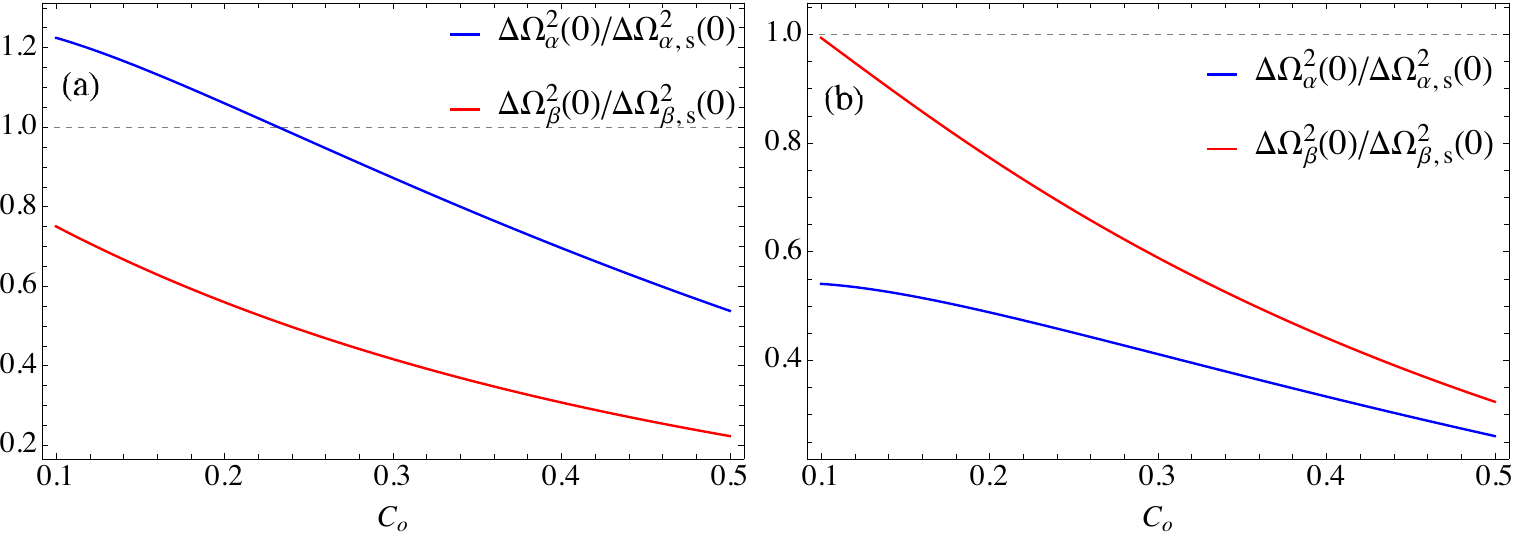}
\caption{(Color online) The comparison of the sensitivities between the multiple-point-coupling system and the traditional system (ii). In (a), the sensitivity $\Delta\Omega_{\varepsilon}^2(0)$ is set at the phase $\phi=\pi$ and selected from the strict braided topology (i). The right nonreciprocity in the multiple-coupling structure can makes both the sensitivities $\Delta\Omega_{\varepsilon}^2(0)$ lower than $\Delta\Omega_{\varepsilon, s}^2(0)$  under suitable cooperativities $C_o$. In (b), the sensitivity $\Delta\Omega_{\varepsilon}^2(0)$ is also set at the phase $\phi=\pi$ but selected from the strict braided topology (ii). Similarly, the left nonreciprocity improves both the sensitivities $\Delta\Omega_{\varepsilon}^2(0)$. }
\label{ComparisonSensitivity2}
\end{figure*}

\noindent\textbf{Traditional System (ii)}: The two modes are directly coupled with interaction Hamiltonian $H_g\rightarrow H_s=i\frac{\sqrt{C_o}\kappa}{2}(a^\dag b_x-H.c.)$.  Donimated by this interaction Hamiltonian, the SNRs in this structure read
\begin{eqnarray}
&&\mathcal{R}_{\alpha, s} (0)=2|1-\frac{(1+\sqrt{C_o})\kappa+\gamma_x+\frac{4\Omega^2}{\gamma_y}}{(1-C_o)\frac{\kappa}{2}+\frac{\gamma_x}{2}+\frac{2\Omega^2}{\gamma_y}}|^2\\
&&\mathcal{R}_{\beta, s} (0)=2|1-\frac{(1-\sqrt{C_o})\kappa}{(1-C_o)\frac{\kappa}{2}+\frac{\gamma_x}{2}+\frac{2\Omega^2}{\gamma_y}}|^2.
\end{eqnarray}
Under the same parameters we used previously, only the output signal $\mathcal{S}_{\alpha, s} (0)$ in this structure can be utilized for readout, as only the SNR $\mathcal{R}_{\alpha, s} (0)$ is greater than 1, as shown in Fig. \ref{ComparisonSNR2}. By further comparing Fig. \ref{ComparisonSNR2} with Fig. \ref{SNR1}, we observe that the SNR $\mathcal{R}_{\alpha, s} (0)\geq\mathcal{R}_{\alpha} (0)$ in both topologies. This result shows that the nonreciprocity in the multiple-point-coupling structure renders the otherwise unreadable output signal $\mathcal{S}_{\beta, s} (0)$ readable; however, this comes at the expense of reducing the output signal $\mathcal{S}_{\alpha, s} (0)$.

Correspondingly, the sensitivity in this sturcture takes the form
\begin{eqnarray}
&&\Delta\Omega^2_{\alpha, s}(0)\simeq\frac{\gamma_y\kappa}{16|\alpha|}|\frac{(1+C_o)^2}{C_o-\sqrt{C_o}}|\\
&&\Delta\Omega^2_{\beta, s}(0)\simeq\frac{\gamma_y\kappa}{16|\alpha|}|\frac{(1+C_o)^2}{1-\sqrt{C_o}}|.
\end{eqnarray}
In Fig. \ref{ComparisonSensitivity2}, we also use the sensitivity ratio $\Delta\Omega^2_{\varepsilon}(0)/\Delta\Omega^2_{\varepsilon, s}(0)$ to show the role of nonreciprocity. Similarly, the sensitivity $\Delta\Omega^2_{\varepsilon}(0)$ in Figs. \ref{ComparisonSensitivity2} (a) and (b) is evaluated at the phase $\phi=\pi$ and is derived from the strict braided topologies (i) and (ii), respectively. As illustrated in Fig. \ref{ComparisonSensitivity2}, both sensitivity ratios $\Delta\Omega^2_{\alpha}(0)/\Delta\Omega^2_{\alpha, s}(0)$ and $\Delta\Omega^2_{\beta}(0)/\Delta\Omega^2_{\beta, s}(0)$ can be less than 1, indicating that nonreciprocity enhances the sensitivity of the gyroscope. The maximum improvement is approximately 70$\%$, corresponding to $\Delta\Omega^2_{\beta}(0)/\Delta\Omega^2_{\beta, s}(0)\approx 0.3$ at $C_o=0.5$ in Fig. \ref{ComparisonSensitivity2} (a) or $\Delta\Omega^2_{\alpha}(0)/\Delta\Omega^2_{\alpha, s}(0)\approx 0.3$ $C_o=0.5$ in Fig. \ref{ComparisonSensitivity2} (b).

Based on the results in Figs. \ref{ComparisonSNR}-\ref{ComparisonSensitivity2}, we observe that nonreciprocity positively influences sensing performance. Compared to traditional reciprocal proposals, the nonreciprocity not only enhances the SNRs making previously unreadable signals readable, but also improves sensitivity. The least favorable outcome involves an increase in sensitivity at the expense of SNR.
\section{Experimental Feasibility\label{Sec7}}
After discussing the performances, we now consider the experimental feasibility of our gyroscope to provide practical guidance. In  Fig. \ref{Model} (b),  we have provided a possible implementation for our gyroscope, which can be tested by nowdays SAWs platform \cite{Schuetz2015, Chu2017, Chu2018, Andersson2019, Noguchi2020}. The whole setup consists of three layers and two zones, fabricated by multi-layer lithograph technology. Detailed fabriacation procedure can be found in the Supplemental Materials of Ref. \cite{Chu2017}. Here, we provide a simple procedure. The Layer 1 (gray block in Fig. \ref{Model} (b)) can be sapphire wafers or metal materials, which is used to etch a mircowave  waveguide. An LC resonator is etched on the Layer 1's surface of the left zone, serving as the auxiliary cavity. The right zone consists of the Layers 2 and 3  made of piezoelectrical materials, e.g., AlN and GaAs, which is used to generate SAWs from the electromagnetic wave of the Layer 1. Notably, the Layers 2 and 3 can be different materials to generate SAWs in different directions, but it needs to ensure that the SAWs in both directions have the same resonant frequency. Additionally, two IDTs should be etched into both layers 2 and 3 to form Bragg gratings, thereby creating a double-mode SAW cavity.  The multiple-point coupling between the auxiliary cavity and the $x$ mode of the SAW cavity can be realized by etching multiple IDTs (see Fig. 1 of Ref. \cite{Andersson2019}).

As mentioned previously, our gyroscope exhibits nonreciprocity, and this nonreciprocity depends on the phase$\phi$ and the number of coupling points $N, M$. Once the setup is fabricated, i.e., the number of coupling points is determined, only the phase $\phi$ can be tuned to acquire nonreciprocity.  A simple method is adjusting the resonant frequency of auxiliary cavity $\omega_a$, which relys on a frequency-tunable LC resonator \cite{Gu2017}. In experiments, one can replace the inductance L with a superconducting quantum inteference device (SQUID) to obtain this tunability. Besides, the driving frequency $\omega_d$ should be tuned synchronously to ensure $\omega_d=\omega_a$, as the system is resonantly driven $\Delta_1=\Delta_2=0$.

Based on accessible parameters in nowadays SAW devices \cite{Schuetz2015, Chu2017, Chu2018, Noguchi2020, Hisatomi2023}, e.g., $\gamma \sim \gamma_{x,y} \sim \mathrm{Hz}$ and $\kappa/\gamma \sim 10-100$,  our gyroscope provide a comparable sensitivity  $\sqrt{\Delta\Omega^2_\varepsilon} \sim 0.1-10~\mathrm{rad/s}$ to classical SAW gyroscopes based on Coriolis effects \cite{Kukaev2025}. Nonetheless, the performance can be further enhanced to exhibit a noticeable advantage, when the quality factor of SAW cavity $Q$ is improved to $10^{10}-10^{12}$.  With this ultra high $Q$,  a SAW cavity operating at GHz has a thermal decay rate $\gamma$ lower than $10^{-3}$ Hz. As a result,  the highest sensitivity of our gyroscope can reach $10^{-5}-10^{-3} rad/s$, enabling us to measure the rotation slower than  the hour hand of a clock, such as self-rotating of the Earth. This is a worthy expected $Q$ factor in near future. A quantum SAW cavity only has a $Q\sim 10^{5}$ when it was first investigated in 2012 \cite{Gustafsson2012}, while it has reached $10^8$ in 2017 \cite{Chu2017}.

\section{Conclusion and outlook \label{Sec8}}
{In conclusion, we present a quantum gyroscope based on coupled giant cavities, where the double-mode SAW cavity serves as the core component forming the fundamental framework of the gyroscope in the $x-y$ plane. Unlike classical SAW gyroscopes, the SAWs used in our proposal are quantized; thus, they transport phonons rather than electrons or holes. This property provides two key advantages: (1) it does not require a dominant Coriolis force to counteract the effects of centrifugal forces, and (2) it enables exploration of the quantum limits (specifically, the shot noise considered in this work) at extremely low pump power, i.e., the sensitivity limits permitted by the Heisenberg uncertainty relation. Following a standard quantum Heisenberg-Langevin equation, we conduct a comprehensive analysis of the system's time-delayed dynamics, which we identify as a quintessential non-Markovian effect. It is important to note that this non-Markovian behavior is an inherent property of quantum giant systems and renders the Markovian approximation inapplicable, regardless of the magnitude of the distances between coupling points. Another consequence of this non-Markovian effect is the induction of topology-dependent nonreciprocal transfer. Although the topologies of coupled giant systems have been studied in previous works \cite{Kockum2018, Kannan2020, Feng2021, Zhu2022B}, the nonreciprocity underlying these systems is discovered here for the first time. Furthermore, we extend the topologies from the two coupling points considered in existing works \cite{Kockum2018, Kannan2020, Feng2021, Zhu2022B} to cases with a greater number of coupling points. This extension allows us to elucidate the role played by the number of coupling points: it complicates the interference effects within the system, thereby altering the conditions for reciprocity. Our further findings regarding sensing performance clarify the role of nonreciprocity in sensing, as it facilitates a more concentrated transfer of energy in one direction. Consequently, the output signal in that direction is not only amplified but also exhibits heightened sensitivity to changes in angular velocity. Compared to traditional reciprocal designs, we demonstrate that nonreciprocity significantly enhances sensor performance, notably improving the SNR, which renders previously unreadable signals discernible and enhances the sensitivity. Under identical parameters, nonreciprocity can enhance sensitivity by approximately $70\%$. The least favorable scenario involves a trade-off, where sensitivity is increased at the expense of SNR.}

{Limited by lithograph technology, nowadays SAW cavities operating at GHz only has a quality factor $Q\sim10^8$, resulting a high thermal decay. This makes our gyroscope  only  provide a comparable sensitivity $0.1-10$ rad/s to the state-of-the-art classical SAW gyroscopes. Nevertheless, we believe that future developments in SAW technology will significantly enhance its quality factor, enabling our multi-point coupling SAW systems to achieve noticeable advantages in future gyroscopes. }

{Since the measurement proposal in this work is based on the spectral height of the output signal, we mainly consider the system's steady-state behavior in the frequency domain. For future studies, one can focus on the non-Markovian time-delayed effect itself in coupled giant systems. This would benefit to clarify how non-Markovianity affect sensing capabilities, and more importantly, it would be helpful to facilitate a real-time control for future quantum coupled-giant systems. }

\section*{Acknowledgements}
 This work is supported by the National Natural Science Foundation of China (NSFC) under Grants No. 62273226, No. 61873162.

\appendix

\section{Derivations of the Hamiltonian, equations of motion, and the transfer function  \label{AppendixA}}
\subsection{Hamiltonian of the double-mode SAW cavity in a rotating coordinate system}
In this section, we begin by introducing the Lagrangian and then proceed to derive the Hamiltonian of a double-mode SAW cavity in a rotating frame, following the standard procedure outlined in our previous work \cite{Zhu2024}. The coordinate systems before and after rotation are illustrated within the red box in Fig. \ref{Model}(a). The $x_o-y_o$ coordinate system (inertial system, plotted in black) rotates counterclockwise with an angular velocity $\Omega$ and then transforms into the $x-y$ coordinate system (non-inertial system, plotted in red). After a time $t$, it has rotated by an angle $\theta=\int_0^t\mathrm{d}\tau~\Omega$. In doing so, the positions of the SAW cavity (shaded box) in the two coordinate systems are $(x_o,y_o)$ and $(x,y)$, respectively. These two coordinate systems are related by the transformation
\begin{eqnarray}
\begin{pmatrix}
x\\
y
\end{pmatrix}
=\begin{pmatrix}
\cos\theta(t) & \sin\theta(t)\\
-\sin\theta(t) & \cos \theta(t)
\end{pmatrix}
\begin{pmatrix}
x_o\\
y_o
\end{pmatrix}.
\label{TransformationRelation}
\end{eqnarray}
The Lagrangian of the oscillator in the original system reads
\begin{equation}
\mathcal{L}_o=T_o-V_o\label{LagrangianO}
\end{equation}
with the kinetic energy $T_o$
\begin{equation}
T_o(\dot{x}_o,\dot{y}_o)=\frac{m}{2} (\dot{x}^2_o+\dot{y}^2_o)\nonumber
\end{equation}
and the potential energy $V_o$
\begin{eqnarray}
&&V_o(x_o,y_o)=\frac{1}{2}[k_x(x-x_e)^2+k_y(y-y_e)^2]\nonumber\\
&&=\frac{k_x}{2}[(x_o-x_{oe})\cos\theta(t)+(y_o-y_{oe})\sin\theta(t)]^2\nonumber\\
&&+\frac{k_y}{2}[-(x_o-x_{oe})\sin\theta(t)+(y_o-y_{oe})\cos\theta(t)]^2,\nonumber
\end{eqnarray}
where $k_{x(y)}$ is the spring constant of the $x(y)$ mode. Here, $x_e, y_e$ and $x_{oe}, y_{oe}$ are the equilibrium positions in the rotation system, while $x_{oe}$ and $y_{oe}$ represent the equilibrium positions in the original system. The Eq. \eqref{LagrangianO} further yields the Hamiltonian
\begin{eqnarray}
&&H_o(x_o,y_o,p_{o,x},p_{o,y})=\frac{1}{2m}(p_{o,x}^2+p_{o,y}^2)\nonumber\\
&&+\frac{k_x}{2}[(x_o-x_{oe})\cos\theta(t)+(y_o-y_{oe})\sin\theta(t)]^2\nonumber\\
&&+\frac{k_y}{2}[-(x_o-x_{oe})\sin\theta(t)+(y_o-y_{oe})\cos\theta(t)]^2,
\end{eqnarray}
where the momenta $p_{x,o}, p_{y,o}$ conjugate to coordinates $x_o,y_o$ read
\begin{eqnarray}
&&p_{x,o}=\frac{\partial \mathcal{L}_o}{\partial \dot{x}_o}=m\dot{x}_o,\label{PXO}\\
&&p_{y,o}=\frac{\partial \mathcal{L}}{\partial \dot{y}_o}=m\dot{y}_o.\label{PYO}
\end{eqnarray}
These operators satisfy the basic commutation relation $[x_o,p_{x,o}]=[y_o,p_{y,o}]=i\hbar$.

In the rotating system, the Lagrangian of the SAW cavity is given by
\begin{eqnarray}
\mathcal{L}=&&T(\dot{x},\dot{y})-V(x,y)\nonumber\\
=&&\frac{m}{2}(\dot{x}^2+\dot{y}^2)-\frac{1}{2}[k_x(x-x_o)^2+k_y(y-y_o)^2]\nonumber\\
&&+m\Omega(-\dot{x}y+x\dot{y})+\frac{m\Omega^2}{2}(x^2+y^2),
\label{LagrangianR}
\end{eqnarray}
where the first term in the second line represents the translational kinetic energy, the second term is the potential energy, the third term denotes the energy induced by Coriolis forces, and the last term is the energy induced by the centrifugal forces. The last two terms are fictitious energy that arises from non-inertial forces. Notably, the Coriolis force does not always dominate  in many complex scenario, such as a high-speed flying aircraft around the Earth. So, one should keep the term of centrifugal force to maintain integrity.

Similarly, this Lagrangian \eqref{LagrangianR} gives the Hamiltonian
\begin{eqnarray}
H(x,y,p_x,p_y)=&&\frac{1}{2m}[(p_x+m\Omega y)^2+(p_y-m\Omega x)^2]\nonumber\\
&&+\frac{1}{2}[k_x(x-x_o)^2+k_y(y-y_o)^2]\nonumber\\
&&-\frac{m\Omega^2}{2}(x^2+y^2),
\label{HamiltonianR}
\end{eqnarray}
where the momenta $p_x,p_y$ conjugate to the coordinates $x,y$ are given by
\begin{subequations}
\begin{eqnarray}
&&p_x=\frac{\partial \mathcal{L}}{\partial \dot{x}}=m\dot{x}-m\Omega y,\label{PXR}\\
&&p_y=\frac{\partial \mathcal{L}}{\partial \dot{y}}=m\dot{y}+m\Omega x\label{PYR}.
\end{eqnarray}
\end{subequations}
The first term in Eq. \eqref{HamiltonianR} is the kinetic energy in the non-inertial system, the second term denotes the potential energy, and the third term represents the energy induced by centrifugal forces.

Furthermore, based on Eqs. \eqref{PXR}-\eqref{PYR} and the relation
\begin{eqnarray}
\begin{pmatrix}
\dot{x}\\
\dot{y}
\end{pmatrix}
=\Omega
\begin{pmatrix}
-\sin\theta(t) & \cos\theta(t)\\
-\cos\theta(t) & -\sin \theta(t)
\end{pmatrix}
\begin{pmatrix}
x_o\\
y_o
\end{pmatrix}\nonumber\\
+\begin{pmatrix}
\cos\theta(t) & \sin\theta(t)\\
-\sin\theta(t) & \cos \theta(t)
\end{pmatrix}
\begin{pmatrix}
\dot{x}_o\\
\dot{y}_o
\end{pmatrix}
\end{eqnarray}
one can examine that the operators $x(y), p_{x(y)}$ also satisfy the basic commutation relation $[x,p_x]=[y,p_y]=i\hbar$.

Using the creation and annihilation operators
\begin{eqnarray}
&&x=\sqrt{\frac{\hbar}{2m\omega_x}}(b_x+b_x^\dag),~~~~p_x=-i\sqrt{\frac{m\hbar\omega_x}{2}}(b_x-b_x^\dag),\nonumber\\
&&y=\sqrt{\frac{\hbar}{2m\omega_y}}(b_y+b_y^\dag),~~~~p_y=-i\sqrt{\frac{m\hbar\omega_y}{2}}(b_y-b_y^\dag),\nonumber
\end{eqnarray}
we can rewrite the Hamiltonian \eqref{HamiltonianR} as a
\begin{eqnarray}
H_m&=&\hbar\omega_x(b_x^\dag b_x+\frac{1}{2})+\hbar\omega_y(b_y^\dag b_y+\frac{1}{2})\nonumber\\
&&+\frac{i\hbar\Omega}{2}[\eta_1(b_x^\dag b_y-b_x b_y^\dag)+\eta_2(b_x^\dag b_y^\dag-b_x b_y)]\nonumber\\
\end{eqnarray}
with the coefficients $\eta_1=\sqrt\frac{\omega_x}{\omega_y}+\sqrt\frac{\omega_y}{\omega_x}$ and $\eta_2=\sqrt\frac{\omega_x}{\omega_y}-\sqrt\frac{\omega_y}{\omega_x}$,
where the resonant frequency is $\omega_{x(y)}=\sqrt{k_{x(y)}/m}$. In the main text, we have assumed $\omega_b=\omega_x=\omega_y$ in order to eliminate the counter-rotating-wave term. For brevity, we let $\hbar=1$ and  set the equilibrium positions $x_o=y_o=0$ for brevity in deriving Eq. \eqref{HM}.  Under these conditions, the Hamiltonian Eq. \eqref{HM} reduces to the second and third terms in $H_{gyro}$ of Eq. \eqref{Hamiltonian}.
\begin{widetext}
\subsection{Derivation of the equations of motion \eqref{Eom}}
The equations of motion \eqref{Eom} is derived from a quantum Heisenberg-Langevin equation, where the operators' dynamics read
\begin{subequations}
\begin{eqnarray}
&&\dot{a}(t)=(-i\omega_a-\frac{\kappa_a}{2})-\sqrt{\kappa_a}\alpha_\mathrm{in}(t)e^{-i\omega_dt}-\sum_{n=1}^N\sqrt{\frac{\gamma_n}{2\pi}}\int\mathrm{d}\omega~c_\omega(t) e^{i\omega\tau_n},\label{EomA}\\
&&\dot{b}_x(t)=(-i\omega_b-\frac{\kappa_b}{2}-\frac{\gamma_x}{2})b_x(t)+\Omega b_y(t)-\sum_{m=1}^M\sqrt{\frac{\bar{\gamma}_m}{2\pi}}\int\mathrm{d}\omega~c_\omega(t) e^{i\omega\bar{\tau}_m}-\sqrt{\kappa_b}\beta_{\mathrm{in}}(t)-\sqrt{\gamma_x}f_x(t),\label{EomBX}\\
&&\dot{b}_y(t)=(-i\omega_b-\frac{\gamma_y}{2})b_y(t)-\Omega b_x(t)-\sqrt{\gamma_y}f_y(t),\label{EomBY}
\end{eqnarray}
\end{subequations}
and
\begin{equation}
\dot{c}_{\omega}(t)=-i\omega c_\omega(t)+\sum_{n=1}^N\sqrt{\frac{\gamma_n}{2\pi}}e^{-i\omega \tau_n} a(t)+\sum_{m=1}^M\sqrt{\frac{\bar{\gamma}_m}{2\pi}}e^{-i\omega \bar{\tau}_m} b_x(t).
\label{EomC}
\end{equation}
The Eq. \eqref{EomC} can be formally solved as
\begin{equation}
c_\omega(t)=c_\omega(0)e^{-i\omega t} +\sum_{n=1}^N\sqrt{\frac{\gamma_n}{2\pi}}\int_{0}^t\mathrm{d} t' a(t')e^{i\omega (t'-t-\tau_n)}+\sum_{m=1}^M\sqrt{\frac{\bar{\gamma}_m}{2\pi}}int_{0}^t\mathrm{d} t' b_x(t')e^{i\omega (t'-t-\bar{\tau}_m)}.
\end{equation}
Substitue this solution into Eqs. \eqref{EomA} and \eqref{EomBX}, one can obatin the following equations of motion of the system's modes
\begin{subequations}
\begin{eqnarray}
\dot{\tilde{a}}(t)&=&(i\Delta_1-\frac{\kappa_a}{2})-\sqrt{\kappa_a}\tilde{\alpha}_\mathrm{in}(t)-\sum_{n=1}^N\sum_{n'=1}^N\sqrt{\gamma_n\gamma_{n'}}\Theta(\tau_n-\tau_{n'})e^{i\omega_d(\tau_n-\tau_{n'})}\tilde{a}(t-(\tau_n-\tau_{n'}))\nonumber\\
&&-\sum_{n=1}^N\sum_{m=1}^M\sqrt{\gamma_n\bar{\gamma}_m}\Theta(\tau_n-\bar{\tau}_{m})e^{i\omega_d(\tau_n-\bar{\tau}_{m})}b_x(t-(\tau_n-\bar{\tau}_{m}))-\sum_{n=1}^N\sqrt{\gamma_n}e^{i\omega_d}\tilde{c}_{\mathrm{in}}(t-\tau_n),\label{EOMReA}\\
\dot{\tilde{b}}_x(t)&=&(i\Delta_2-\frac{\kappa_b}{2}-\frac{\gamma_x}{2})\tilde{b}_x(t)+\Omega b_y(t)-\sqrt{\kappa_b}\tilde{\beta}_\mathrm{in}(t)-\sqrt{\gamma_y}\tilde{f}_y(t)\nonumber\\
&&-\sum_{m=1}^M\sum_{m'=1}^M\sqrt{\bar{\gamma}_m\bar{\gamma}_{m'}}\Theta(\bar{\tau}_m-\bar{\tau}_{m'})e^{i\omega_d(\bar{\tau}_n-\bar{\tau}_{m'})}\tilde{b}_x(t-(\bar{\tau}_m-\bar{\tau}_{m'}))\nonumber\\
&&-\sum_{n=1}^N\sum_{m=1}^M\sqrt{\gamma_n\bar{\gamma}_m}\Theta(\bar{\tau}_m-{\tau}_{n})e^{i\omega_d(\bar{\tau}_m-{\tau}_{n})}\tilde{a}(t-(\bar{\tau}_m-{\tau}_{m}))-\sum_{m=1}^M\sqrt{\bar{\gamma}_m}e^{i\omega_d}\tilde{c}_{\mathrm{in}}(t-\bar{\tau}_m)-\sqrt{\gamma_x}\tilde{f}_x(t),\label{EOMReBX}\\
\dot{\tilde{b}}_y(t)&=&(i\Delta_2-\frac{\gamma_y}{2})-\Omega\tilde{b}_x(t)-\sqrt{\gamma_y}\tilde{f}_y(t),\label{EOMReBY}
\end{eqnarray}
\end{subequations}
where the detunings are $\Delta_1=\omega_d-\omega_a$ and $\Delta_2=\Delta_1+\omega_a-\omega_b$.  Here, we use $O(t)=\tilde{O}(t)e^{-i\omega_dt}$ denotes the slowly-varying form of the operator $O$. For simple notions, we still use $O(t)$ without a tilde to lable its slowly-varying form in the main text.
Let $\gamma_n=\bar{\gamma}_m=\gamma$ and reorganize the Eqs. \eqref{EOMReA}-\eqref{EOMReBY} in a matrix form, one can obtain the Eq. \eqref{Eom} in the main text with the coefficient matrices
\begin{subequations}
\begin{eqnarray}
A_o=
\begin{pmatrix}
i\Delta_1-\frac{\kappa_a}{2} & 0 & 0 \\
0 & i\Delta_2-\frac{\kappa_b+\gamma_x}{2} & \Omega\\
0 & -\Omega & i\Delta_2-\frac{ \gamma_y}{2}
\end{pmatrix},~~~
B_o=\begin{pmatrix}
\sqrt{\kappa_a} & 0  & 0 & 0\\
0 &\sqrt{\kappa_b}   &\sqrt{\gamma_x} & 0\\
0 & 0  & 0 &  \sqrt{\gamma_y}
\end{pmatrix}.
\end{eqnarray}
\end{subequations}
\subsection{Detailed form of the transfer function}
In the frequency domain, the matrices $\mathcal{A}(\omega)$ and $\mathcal{B}(\omega)$ read
\begin{equation}
\begin{aligned}
\mathcal{A}(\omega)=\mathcal{A}_o(\omega)+\mathcal{A}_{g}(\omega),~~
\mathcal{B}(\omega)=\mathcal{B}_o(\omega)+\mathcal{B}_{g}(\omega),
\end{aligned}
\label{CMatrices}
\end{equation}
with
\begin{subequations}
\begin{eqnarray}
&&\mathcal{A}_o(\omega)=
\mathrm{diag}~\big(\chi_1(\omega), \chi_2(\omega)\big),~~~~~~~~~~~~~~~~~~~~~
\mathcal{A}_{g}(\omega)=
\begin{pmatrix}
\Gamma(\omega_d-\omega;\tau_n,\tau_{n'}) & \Gamma(\omega_d-\omega; \tau_n,\bar{\tau}_{m}) \\
\Gamma(\omega_d-\omega;\bar{\tau}_{m},\tau_n) & \Gamma(\omega_d-\omega;\bar{\tau}_m,\bar{\tau}_{m'})
\end{pmatrix},\nonumber\\
&&\mathcal{B}_o=
\begin{pmatrix}
\sqrt{\kappa_a} & 0 & 0 & 0 & 0 \\
0 & \sqrt{\kappa_b} & 0 &\sqrt{\gamma_x} & -\frac{\Omega\sqrt{\gamma_y}}{i(-\omega+\Delta_2)-\frac{\gamma_y}{2}}
\end{pmatrix},~~
\mathcal{B}_{g}(\omega)=\begin{pmatrix}
0 & 0 & \sum_{n=1}^N\sqrt{\gamma}e^{i(\omega_d-\omega)\tau_n} & 0 & 0  \\
0 & 0 & \sum_{m=1}^M\sqrt{\gamma}e^{i(\omega_d-\omega)\bar{\tau}_m} & 0 & 0
\end{pmatrix}\nonumber,
\end{eqnarray}
\end{subequations}
where susceptibility functions are $\chi_1(\omega)=i(-\omega+\Delta_1)-\frac{\kappa_a}{2}$ and $\chi_2(\omega)= i(-\omega+\Delta_2)-\frac{\kappa_b+\gamma_x}{2}+\frac{\Omega^2}{i(-\omega+\Delta_2)-\frac{\gamma_y}{2}}$.

According to Eq. \eqref{TransferFunction}, the general form of transfer function reads
\begin{eqnarray}
G(\omega)=\begin{pmatrix}
1 & 0 & 0	& 0 & 0\\
0 & 1 & 0	& 0 & 0
\end{pmatrix}+\frac{1}{\mathrm{det}~\mathcal{A}(\omega)}
\begin{pmatrix}
\kappa_a[\chi_2(\omega)+\Gamma(\omega_d-\omega;\bar{\tau}_m,\bar{\tau}_{m'})] & -\sqrt{\kappa_a\kappa_b}~\Gamma(\omega_d-\omega;\tau_n,\bar{\tau}_m) & \#_{13} & \#_{14} & \#_{15}\\
-\sqrt{\kappa_a\kappa_b}~\Gamma(\omega_d-\omega;\bar{\tau}_m,\tau_n) & \kappa_b[\chi_1(\omega)+\Gamma(\omega_d-\omega;{\tau}_n,\tau_{n'})] & \#_{23} & \#_{24} & \#_{25}
\end{pmatrix}\nonumber
\end{eqnarray}
with
\begin{eqnarray}
&&\mathrm{det}~\mathcal{A}(\omega)=[\chi_1(\omega)+\Gamma(\omega_d-\omega;{\tau}_n,\tau_{n'})][\chi_2(\omega)+\Gamma(\omega_d-\omega;\bar{\tau}_m,\bar{\tau}_{m'})]-\Gamma(\omega_d-\omega;\tau_n,\bar{\tau}_m)\Gamma(\omega_d-\omega;\bar{\tau}_m,\tau_n)\nonumber\\
&&\#_{13}=\sqrt{\kappa_a\gamma}~\{\sum_{n=1}^N[\chi_2(\omega)+\Gamma(\omega_d-\omega;\bar{\tau}_m,\bar{\tau}_{m'})]e^{i(\omega_d-\omega)\tau_n}-\sum_{m=1}^M\Gamma(\omega_d-\omega;\tau_n,\bar{\tau}_{m})e^{i(\omega_d-\omega)\bar{\tau}_m}\}\nonumber\\
&&\#_{14}=-\sqrt{\kappa_a\gamma_x}~\Gamma(\omega_d-\omega;\tau_n,\bar{\tau}_{m})\nonumber\\
&&\#_{15}=-\sqrt{\kappa_a\gamma_y}~\Gamma(\omega_d-\omega;\tau_n,\bar{\tau}_{m})\frac{\Omega}{i(-\omega+\Delta_2)-\frac{\gamma_y}{2}}\nonumber\\
&&\#_{23}=\sqrt{\kappa_b\gamma}~\{\sum_{m=1}^M[\chi_1(\omega)+\Gamma(\omega_d-\omega;\tau_n,{\tau}_{n'})]e^{i(\omega_d-\omega)\bar{\tau}_m}-\sum_{n=1}^N\Gamma(\omega_d-\omega;\bar{\tau}_{m}, \tau_n)e^{i(\omega_d-\omega){\tau}_n}\}\nonumber\\
&&\#_{24}=\sqrt{\kappa_b\gamma_x}[\chi_1(\omega)+\Gamma(\omega_d-\omega;\tau_n,{\tau}_{n'})]\nonumber\\
&&\#_{25}=\sqrt{\kappa_b\gamma_y}[\chi_1(\omega)+\Gamma(\omega_d-\omega;\tau_n,{\tau}_{n'})]\frac{\Omega}{i(-\omega+\Delta_2)-\frac{\gamma_y}{2}}.\nonumber
\end{eqnarray}
Notably, the output signals $\mathcal{S}_{\epsilon}(\omega), \epsilon=\alpha, \beta$ \eqref{ABRelation} only depend solely on the first two columns of the matrix $G(\omega)$. This explains our exclusive focus on the elements $G_{12}(\omega)$ and $G_{21}(\omega)$ when discussing the nonreciprocal strength $\sigma(\omega)$. Furthermore, since the angular velocity $\Omega$ is incorporated into the susceptibility function $\chi_2(\omega)$, the elements $G_{12}(\omega)$ and $G_{21}(\omega)$ cannot be zero. This condition is essential to ensure that both output signals $\mathcal{S}_{\alpha}(\omega)$ and $\mathcal{S}_{\beta}(\omega)$ contain the angular velocity $\Omega$, which aids in distinguishing between separated and nested topologies.

\section{The coupling matrix $\mathcal{A}_g(\omega)$ and the topology-dependent  nonreciprocal strength $\sigma(\omega)$\label{AppendixB}}
In this Appendix, we provide the detailed form of coupling matrix $\mathcal{A}_g(\omega)$.

\noindent\textbf{(a) Separated.}  In this topology, the matrix $\mathcal{A}_g(\omega)=
\begin{pmatrix}
\mathcal{A}_{g,11}(\omega) & \mathcal{A}_{g,12}(\omega)\\
\mathcal{A}_{g,21}(\omega) & \mathcal{A}_{g,22}(\omega)
\end{pmatrix}$ has the elements
\begin{equation}
\begin{aligned}
&\mathcal{A}_{g,11}(\omega)=-\gamma
\begin{cases}
\frac{N}{2}+\sum_{s=1}^N(N-s)e^{si\phi} & \mathrm{(i)}\\
\frac{N}{2}+\sum_{s=1}^N(N-s)e^{si\phi} & \mathrm{(ii)}
\end{cases},~~
\mathcal{A}_{g,12}(\omega)=-\gamma
\begin{cases}
0 & \mathrm{(i)}\\
\sum_{k=1}^N\sum_{s=k}^{k+M-1}e^{si\phi} &\mathrm{(ii)}
\end{cases},\\
&\mathcal{A}_{g,21}(\omega)=-\gamma
\begin{cases}
\sum_{k=1}^M\sum_{s=k}^{k+N-1}e^{si\phi} 	& \mathrm{(i)}\\
0 & \mathrm{(ii)}
\end{cases},~~~~~~
\mathcal{A}_{g,22}(\omega)=-\gamma
\begin{cases}
\frac{M}{2}+\sum_{s=1}^M(M-s)e^{si\phi} 	& \mathrm{(i)}\\
\frac{M}{2}+\sum_{s=1}^M(M-s)e^{si\phi}	& \mathrm{(ii)}
\end{cases}.\\
\end{aligned}
\label{SeparatedAg}
\end{equation}
For $\phi=\{0,2\pi\}$, the matrix $\mathcal{A}_g(\omega)$ read
\begin{eqnarray}
\mathcal{A}_g(\omega)=-\gamma
\begin{cases}
\begin{pmatrix}
\frac{N^2}{2} & 0 \\
NM	& \frac{M^2}{2}
\end{pmatrix}	& (\mathrm{i})\\
\begin{pmatrix}
\frac{N^2}{2} & NM \\
0	& \frac{M^2}{2}
\end{pmatrix}	& (\mathrm{ii})
\end{cases}.\nonumber
\end{eqnarray}
For $\phi\neq\{0, 2\pi\}$, it becomes
\begin{eqnarray}
\mathcal{A}_g(\omega)=-\gamma
\begin{cases}
\begin{pmatrix}
\frac{N}{2}+\frac{e^{i\phi}[N(1-e^{i\phi})-(1-e^{Ni\phi})]}{(1-e^{i\phi})^2} & 0 \\
\frac{e^{i\phi}(1-e^{Mi\phi})(1-e^{Ni\phi})}{(1-e^{i\phi})^2}	& \frac{M}{2}+\frac{e^{i\phi}[M(1-e^{i\phi})-(1-e^{Mi\phi})]}{(1-e^{i\phi})^2}
\end{pmatrix}	& (\mathrm{i})\\
\begin{pmatrix}
\frac{N}{2}+\frac{e^{i\phi}[N(1-e^{i\phi})-(1-e^{Ni\phi})]}{(1-e^{i\phi})^2} & \frac{e^{i\phi}(1-e^{Mi\phi})(1-e^{Ni\phi})}{(1-e^{i\phi})^2} \\
0	& \frac{M}{2}+\frac{e^{i\phi}[M(1-e^{i\phi})-(1-e^{Mi\phi})]}{(1-e^{i\phi})^2}
\end{pmatrix}	& (\mathrm{ii})
\end{cases}.\nonumber
\end{eqnarray}
In this case, the nonreciprocal strength $\sigma(\omega) $ is consistently $1$ or $-1$ irrespective of the phase $\phi$.

\noindent\textbf{(b) Nested.} Accordingly, the elements of the matrix $\mathcal{A}_{g}(\omega)$ take the form
\begin{equation}
\begin{aligned}
&\mathcal{A}_{g,11}(\omega)=-\gamma
\begin{cases}
\frac{N}{2}+E_1 & \mathrm{(i)}\\
\frac{N}{2}+\sum_{s=1}^N(N-s)e^{si\phi}  & \mathrm{(ii)}
\end{cases},~
\mathcal{A}_{g,12}(\omega)=-\gamma
\begin{cases}
\sum_{k=1}^{N-n}\sum_{s=k}^{k+M-1}e^{si\phi} & \mathrm{(i)}\\
\sum_{k=1}^{N}\sum_{s=k}^{k+m-1}e^{si\phi} & \mathrm{(ii)}
\end{cases},\\
&\mathcal{A}_{g,21}(\omega)=-\gamma
\begin{cases}
\sum_{k=1}^M\sum_{s=k}^{k+n-1} e^{si\phi} & \mathrm{(i)}\\
\sum_{k=1}^{M-m}\sum_{s=k}^{k+N-1} e^{si\phi} & \mathrm{(ii)}
\end{cases},~~
\mathcal{A}_{g,22}(\omega)=-\gamma
\begin{cases}
\frac{M}{2}+\sum_{s=1}^M(M-s)e^{si\phi} & \mathrm{(i)}\\
\frac{M}{2}+E_2 & \mathrm{(ii)}
\end{cases},\\
\end{aligned}
\label{NestedAg}
\end{equation}
with $E_1=\sum_{s=1}^n(n-s)e^{si\phi}+\sum_{s=1}^{N-n}(N-n-s)e^{si\phi}+\sum_{k=1}^{N-n}\sum_{s=k}^{k+n-1}e^{(s+M)i\phi}$ and $E_2=\sum_{s=1}^m(m-s)e^{si\phi}+\sum_{s=1}^{M-m}(M-m-s)e^{si\phi}+\sum_{k=1}^{M-m}\sum_{s=k}^{k+m-1}e^{(s+N)i\phi}$. For $\phi=\{0, 2\pi\}$, the matrix $\mathcal{A}_{g}(\omega)$ reads
\begin{eqnarray}
\mathcal{A}_g(\omega)=-\gamma
\begin{cases}
\begin{pmatrix}
\frac{N(N+1)}{2}+(N-n)^2-n & M(N-n) \\
Mn & \frac{M^2}{2}
\end{pmatrix}	& (\mathrm{i})\\
\begin{pmatrix}
\frac{N^2}{2} & Nm\\
N(M-m) & \frac{M(M+1)}{2}+(M-m)^2+m
\end{pmatrix}	& (\mathrm{ii})
\end{cases}.\nonumber
\end{eqnarray}
For  $\phi\neq\{0, 2\pi\}$, it becomes
\begin{eqnarray}
\mathcal{A}_g(\omega)=-\gamma
\begin{cases}
\begin{pmatrix}
\frac{N}{2}+\frac{e^{i\phi}\{N(1-e^{i\phi})-(1-e^{ni\phi})+[e^{Mi\phi}(1-e^{ni\phi})-1](1-e^{(N-n)i\phi})\}}{(1-e^{i\phi})^2} & \frac{e^{i\phi}(1-e^{(N-n)i\phi})(1-e^{Mi\phi})}{(1-e^{i\phi})^2}\\
\frac{e^{i\phi}(1-e^{n i\phi})(1-e^{Mi\phi})}{(1-e^{i\phi})^2} & \frac{M}{2}+\frac{e^{i\phi}[M(1-e^{i\phi})-(1-e^{Mi\phi})]}{(1-e^{i\phi})^2}
\end{pmatrix}	& (\mathrm{i})\\
\begin{pmatrix}
\frac{N}{2}+\frac{e^{i\phi}[N(1-e^{i\phi})-(1-e^{Ni\phi})]}{(1-e^{i\phi})^2} & \frac{e^{i\phi}(1-e^{N i\phi})(1-e^{mi\phi})}{(1-e^{i\phi})^2}\\
\frac{e^{i\phi}(1-e^{Ni\phi})(1-e^{(M-m)i\phi})}{(1-e^{i\phi})^2} & \frac{M}{2}+\frac{e^{i\phi}\{M(1-e^{i\phi})-(1-e^{mi\phi})+[e^{Ni\phi}(1-e^{mi\phi})-1](1-e^{(M-m)i\phi})\}}{(1-e^{i\phi})^2}
\end{pmatrix}	& (\mathrm{ii})
\end{cases}.\nonumber
\end{eqnarray}
Correspondingly, the nonreciprocal strength $\sigma(\omega)$ is
\begin{equation}
\sigma(\omega)=
\begin{cases}
\begin{cases}
\frac{n^2-(N-n)^2}{n^2-(N-n)^2} & \phi=\{0, 2\pi\}\\
\frac{|1-e^{ni\phi}|^2-|1-e^{(N-n)i\phi}|^2}{|1-e^{ni\phi}|^2+|1-e^{(N-n)i\phi}|^2} &  \phi\neq\{0, 2\pi\}
\end{cases} & (\mathrm{i}) \\
\begin{cases}
\frac{(M-m)^2-m^2}{(M-m)^2+m^2} & \phi=\{0, 2\pi\}\\
\frac{|1-e^{(M-m)i\phi}|^2-|1-e^{mi\phi}|^2}{|1-e^{(M-m)i\phi}|^2+|1-e^{mi\phi}|^2} &   \phi\neq\{0, 2\pi\}
\end{cases} &  (\mathrm{ii})
\end{cases}. \nonumber
\end{equation}
This further yields the condition for reciprocity
\begin{equation}
 \sigma(\omega)=0\Rightarrow\sin\frac{N}{2}\phi\sin\frac{N-2n}{2}\phi=0,
\label{NestedCondition}
\end{equation}
 such that the phase $\phi$ should satisfy
\begin{equation}
\phi=
\begin{cases}
\begin{cases}
arbitrary & N=2n \\
\frac{2\pi}{N} & |N-2n|=1 \\
\{\frac{2\pi}{N}, \frac{2\pi}{|N-2n|}\}, & N\neq2n, |N-2n|\neq1
\end{cases}	& (\mathrm{i})\\
\begin{cases}
arbitrary & M=2m \\
\frac{2\pi}{M} & |M-2m|=1 \\
\{\frac{2\pi}{M}, \frac{2\pi}{|M-2m|}\}, & M\neq2m, |M-2m|\neq1
\end{cases} & (\mathrm{ii})
\end{cases}.\nonumber
\end{equation}
\noindent\textbf{(c) Strict Braided.} Correspondingly, the elements of the matrix $\mathcal{A}^S_{g}(\omega)$ are given by
\begin{equation}
\begin{aligned}
&\mathcal{A}^S_{g,11}(\omega)=-\gamma
\begin{cases}
\frac{N}{2}+\sum_{s=1}^N(N-s)e^{2si\phi} & \mathrm{(i)}\\
\frac{N}{2}+\sum_{s=1}^N(N-s)e^{2si\phi} & \mathrm{(ii)}
\end{cases},~~~~~
\mathcal{A}^S_{g,12}(\omega)=-\gamma
\begin{cases}
\sum_{s=1}^{N-1}(N-s)e^{(2s-1)i\phi} & \mathrm{(i)}\\
\sum_{s=1}^{M}(M+1-s)e^{(2s-1)i\phi} & \mathrm{(ii)}
\end{cases},\\
&\mathcal{A}^S_{g,21}(\omega)=-\gamma
\begin{cases}
\sum_{s=1}^{N}(N+1-s)e^{(2s-1)i\phi} & \mathrm{(i)}\\
\sum_{s=1}^{M-1}(M-s)e^{(2s-1)i\phi} & \mathrm{(ii)}
\end{cases},~
\mathcal{A}^S_{g,22}(\omega)=-\gamma
\begin{cases}
\frac{M}{2}+\sum_{s=1}^M(M-s)e^{2si\phi} & \mathrm{(i)}\\
\frac{M}{2}+\sum_{s=1}^M(M-s)e^{2si\phi} & \mathrm{(ii)}
\end{cases}.
\end{aligned}
\label{StrictBraidedAg}
\end{equation}
For $\phi=\{0, \pi, 2\pi\}$, the matrix $\mathcal{A}^S_{g}(\omega)$ reads
\begin{eqnarray}
\mathcal{A}^S_g(\omega)=-\gamma
\begin{cases}
\begin{pmatrix}
\frac{N^2}{2} & -\frac{N(N-1)}{2} \\
-\frac{N(N+1)}{2} & \frac{M^2}{2}
\end{pmatrix}	& (\mathrm{i})\\
\begin{pmatrix}
\frac{N^2}{2} & -\frac{M(M+1)}{2}\\
-\frac{M(M-1)}{2} & \frac{M^2}{2}
\end{pmatrix}	& (\mathrm{ii})
\end{cases}.\nonumber
\end{eqnarray}
For $\phi\neq\{0, \pi, 2\pi\}$, it takes the form
\begin{eqnarray}
\mathcal{A}^S_g(\omega)=-\gamma
\begin{cases}
\begin{pmatrix}
\frac{N}{2}+\frac{e^{2i\phi}[N(1-e^{2i\phi})-(1-e^{2Ni\phi})]}{(1-e^{2i\phi})^2} & \frac{e^{i\phi}[(N-1)(1-e^{2i\phi})-e^{2i\phi}(1-e^{2(N-1)i\phi})]}{(1-e^{2i\phi})^2} \\
\frac{e^{i\phi}[N(1-e^{2i\phi})-e^{2i\phi}(1-e^{2Ni\phi})]}{(1-e^{2i\phi})^2} & \frac{M}{2}+\frac{e^{2i\phi}[M(1-e^{2i\phi})-(1-e^{2Mi\phi})]}{(1-e^{2i\phi})^2}
\end{pmatrix}	& (\mathrm{i})\\
\begin{pmatrix}
\frac{N}{2}+\frac{e^{2i\phi}[N(1-e^{2i\phi})-(1-e^{2Ni\phi})]}{(1-e^{2i\phi})^2} & \frac{e^{i\phi}[M(1-e^{2i\phi})-e^{2i\phi}(1-e^{2Mi\phi})]}{(1-e^{2i\phi})^2} \\
\frac{e^{i\phi}[(M-1)(1-e^{2i\phi})-e^{2i\phi}(1-e^{2(M-1)i\phi})]}{(1-e^{2i\phi})^2} & \frac{M}{2}+\frac{e^{2i\phi}[M(1-e^{2i\phi})-(1-e^{2Mi\phi})]}{(1-e^{2i\phi})^2}
\end{pmatrix}	& (\mathrm{ii})
\end{cases}.\nonumber
\end{eqnarray}

The corresponding nonreciprocal strength $\sigma^S(\omega)$ takes the form
\begin{equation}
\sigma^S(\omega)=
\begin{cases}
\begin{cases}
\frac{(N+1)^2-(N-1)^2}{(N+1)^2+(N-1)^2}>0 & \phi=\{0, \pi, 2\pi\}\\
\frac{|N(1-e^{2i\phi})-e^{2i\phi}(1-e^{2Ni\phi})|^2-|(N-1)(1-e^{2i\phi})-e^{2i\phi}(1-e^{2(N-1)i\phi})|^2}{|N(1-e^{2i\phi})-e^{2i\phi}(1-e^{2Ni\phi})|^2+|(N-1)(1-e^{2i\phi})-e^{2i\phi}(1-e^{2(N-1)i\phi})|^2} &  \phi\neq\{0, \pi ,2\pi\}
\end{cases} & (\mathrm{i}) \\
\begin{cases}
\frac{(M-1)^2-(M+1)^2}{(M-1)^2+(M+1)^2}<0 & \phi=\{0,\pi, 2\pi\}\\
\frac{|(M-1)(1-e^{2i\phi})-e^{2i\phi}(1-e^{2(M-1)i\phi})|^2-|M(1-e^{2i\phi})-e^{2i\phi}(1-e^{2Mi\phi})|^2}{|(M-1)(1-e^{2i\phi})-e^{2i\phi}(1-e^{2(M-1)i\phi})|^2+|M(1-e^{2i\phi})-e^{2i\phi}(1-e^{2Mi\phi})|^2} &   \phi\neq\{0,\pi, 2\pi\}
\end{cases} &  (\mathrm{ii})
\end{cases}.\nonumber
\end{equation}
In this case, the reciprocal points
\begin{equation}
\sigma^S(\omega)=0\Rightarrow\sin^2\phi\sin^2N\phi=0
\label{StrictCondtion}
\end{equation}
only occurs at the phase
\begin{eqnarray}
\phi=
\begin{cases}
\{\frac{\pi}{N}, \frac{\pi}{N}+\pi\} & (\mathrm{i}) \\
\{\frac{\pi}{M}, \frac{\pi}{M}+\pi\} & (\mathrm{ii})
\end{cases}.\nonumber
\end{eqnarray}
One can observe that the nonreciprocal strength $\sigma(\omega)$ is influenced by both the topologies and the number of coupling points. Thus, even with the same topology, varying the number of coupling points alters the nonreciprocity of the system.
\section{Discussions on the general braided topology \label{AppendixC}}

\subsection{The general form of matrix $\mathcal{A}^G_{g}(\omega)$ and the associated nonreciprocal strength $\sigma^G(\omega)$}
In the general braided topology, the matrix $\mathcal{A}^G_{g}(\omega)$ reads
\begin{eqnarray}
\mathcal{A}^G_{g}(\omega)=\mathcal{A}^S_{g}(\omega)+\mathcal{A}^A_{g}(\omega), \label{Type3G}
\end{eqnarray}
where the matrix $\mathcal{A}^A_{g}(\omega)$ results from the addtional seperated components, c.f. Fig. \ref{EffectiveTop} (c). Here, the matrix $\mathcal{A}^A_{g}(\omega)$ has the elements
\begin{eqnarray}
&&\mathcal{A}^A_{g,11}(\omega)=-\gamma
\begin{cases}
0 & \mathrm{(i)}\\
\sum_{s=1}^{N-M}(N-M-s)e^{2si\phi}+\sum_{p=1}^{\lfloor\frac{N-M}{2}\rfloor}\sum_{s=p}^{p+M-1}e^{2si\phi}+\sum_{q=1}^{\lceil\frac{N-M}{2}\rceil}\sum_{s=q}^{q+M-1}e^{(2s-1)i\phi}] & \mathrm{(ii)}
\end{cases}\nonumber\\
&&\mathcal{A}^A_{g,12}(\omega)=-\gamma
\begin{cases}
0 & \mathrm{(i)}\\
\sum_{p=1}^{\lceil\frac{N-M}{2}\rceil}\sum_{s=p}^{p+M-1}e^{2si\phi}+\sum_{q=1}^{\lfloor\frac{N-M}{2}\rfloor}\sum_{s=q}^{q+M-1}e^{2(s+1)i\phi} & \mathrm{(ii)}
\end{cases}\nonumber\\
&&\mathcal{A}^A_{g,21}(\omega)=-\gamma
\begin{cases}
\sum_{p=1}^{\lceil\frac{M-N}{2}\rceil}\sum_{s=p}^{p+N-1}e^{2si\phi}+\sum_{q=1}^{\lfloor\frac{M-N}{2}\rfloor}\sum_{s=q}^{q+N-1}e^{(2s+1)i\phi} & \mathrm{(i)}\\
0 & \mathrm{(ii)}
\end{cases}\nonumber\\
&&\mathcal{A}^A_{g,22}(\omega)=-\gamma
\begin{cases}
\sum_{s=1}^{M-N}(M-N-s)e^{2si\phi}+\sum_{p=1}^{\lfloor\frac{M-N}{2}\rfloor}\sum_{s=p}^{p+N-1}e^{2si\phi}+\sum_{q=1}^{\lceil\frac{M-N}{2}\rceil}\sum_{s=q}^{q+N-1}e^{(2s-1)i\phi}] & \mathrm{(i)}\\
0 & \mathrm{(ii)}
\end{cases}\nonumber
\end{eqnarray}
where the ceiling (flooring) functions is $\lceil\rceil$ ($\lfloor\rfloor$). If $M-N=2k, k\in \mathrm{Z}^+$, $\lceil(M-N)/2\rceil=\lfloor(M-N)/2\rfloor$. Notably, the general braided topology (i) requires $M-N\geq1, N\geq2$ while its mirror image (ii) requires $N-M\geq1, M\geq2$.

For $\phi=\{0,\pi, 2\pi\}$, the matrix $\mathcal{A}^G_g(\omega)$ in the general braided topologies (i) and (ii) read
\begin{eqnarray}
\mathcal{A}^G_g(\omega)=-\gamma
\begin{cases}
\begin{pmatrix}
\frac{N^2}{2} & -\frac{N(N-1)}{2} \\
-\frac{N(N-1)}{2} & \frac{(M-N)(M-N-1)+N(N-1)+M}{2}
\end{pmatrix}	& M-N=2k-1\\
\begin{pmatrix}
\frac{N^2}{2} & -\frac{N(N-1)}{2} \\
-\frac{N(N+1)}{2} & \frac{(M-N)(M-N-1)+N(N+1)+M}{2}
\end{pmatrix} & M-N=2k\\
\end{cases} & (\mathrm{i}),
\end{eqnarray}
and
\begin{eqnarray}
\mathcal{A}^G_g(\omega)=-\gamma
\begin{cases}
\begin{pmatrix}
\frac{(N-M)(N-M-1)+M(M-1)+N}{2} & -\frac{M(M-1)}{2} \\
-\frac{M(M-1)}{2} & \frac{M^2}{2}
\end{pmatrix}	& N-M=2k-1\\
\begin{pmatrix}
\frac{(N-M)(N-M-1)+M(M+1)+N}{2} & -\frac{M(M+1)}{2} \\
-\frac{M(M-1)}{2} & \frac{M^2}{2}
\end{pmatrix}	& N-M=2k
\end{cases} & ~(\mathrm{ii}).
\end{eqnarray}
Correspondingly, the nonreciprocal strength $\sigma(\omega)$ is
\begin{eqnarray}
\sigma(\omega)=
\begin{cases}
\begin{cases}
0 & M-N=2k-1\\
\frac{(N+1)^2-(N-1)^2}{(N+1)^2+(N-1)^2} &  M-N=2k
\end{cases} & (\mathrm{i}) \\
\begin{cases}
0 & N-M=2k-1\\
\frac{(M-1)^2-(M+1)^2}{(M-1)^2+(M+1)^2} &   N-M=2k
\end{cases} &  (\mathrm{ii})
\end{cases}.
\end{eqnarray}
For $\phi\neq\{0,\pi, 2\pi\}$, the matrix $\mathcal{A}^G_g(\omega)$ becomes
\begin{eqnarray}
\mathcal{A}^G_g(\omega)=-\gamma
\begin{cases}
\begin{pmatrix}
\frac{N}{2}+\frac{e^{2i\phi}[N(1-e^{2i\phi})-(1-e^{2Ni\phi})]}{(1-e^{2i\phi})^2} & \frac{e^{i\phi}[(N-1)(1-e^{2i\phi})-e^{2i\phi}(1-e^{2(N-1)i\phi})]}{(1-e^{2i\phi})^2} \\
\frac{e^{i\phi}[N(1-e^{2i\phi})+e^{i\phi}(1-e^{2Ni\phi})(1-e^{(M-N)i\phi}-e^{(M-N+1)i\phi})]}{(1-e^{2i\phi})^2} & \frac{M}{2}+\Phi_1
\end{pmatrix}	& (\mathrm{i})\\
\begin{pmatrix}
\frac{N}{2}+\Phi_2 & \frac{e^{i\phi}[M(1-e^{2i\phi})+e^{i\phi}(1-e^{2Mi\phi})(1-e^{(N-M)i\phi}-e^{(N-M+1)i\phi})]}{(1-e^{2i\phi})^2} \\
\frac{e^{i\phi}[(M-1)(1-e^{2i\phi})-e^{2i\phi}(1-e^{2(M-1)i\phi})]}{(1-e^{2i\phi})^2} & \frac{M}{2}+\frac{e^{2i\phi}[M(1-e^{2i\phi})-(1-e^{2Mi\phi})]}{(1-e^{2i\phi})^2}
\end{pmatrix} & (\mathrm{ii})
\end{cases}
\end{eqnarray}
with
\begin{eqnarray}
&&\Phi_1=\frac{e^{2i\phi}[M(1-e^{2i\phi})-(1-e^{2(M-N)i\phi})+(1-e^{2Ni\phi})(e^{-i\phi}-e^{(M-N-1)i\phi}-e^{(M-N)i\phi})]}{(1-e^{2i\phi})^2}\nonumber\\
&&\Phi_2=\frac{e^{2i\phi}[N(1-e^{2i\phi})-(1-e^{2(N-M)i\phi})+(1-e^{2Mi\phi})(e^{-i\phi}-e^{(N-M-1)i\phi}-e^{(N-M)i\phi})]}{(1-e^{2i\phi})^2}.\nonumber
\end{eqnarray}
Accordingly, the nonreciprocal strength $\sigma^G(\omega)$ reads
\begin{eqnarray}
\sigma^G(\omega)=
\begin{cases}
\frac{|N(1-e^{2i\phi})+e^{i\phi}(1-e^{2Ni\phi})(1-e^{(M-N)i\phi}-e^{(M-N+1)i\phi})|^2-|(N-1)(1-e^{2i\phi})-e^{2i\phi}(1-e^{2(N-1)i\phi})|^2}{N(1-e^{2i\phi})+e^{i\phi}(1-e^{2Ni\phi})(1-e^{(M-N)i\phi}-e^{(M-N+1)i\phi})|^2+|(N-1)(1-e^{2i\phi})-e^{2i\phi}(1-e^{2(N-1)i\phi})|^2} &  (\mathrm{i})\\
\frac{|(M-1)(1-e^{2i\phi})-e^{2i\phi}(1-e^{2(M-1)i\phi})|^2-|M(1-e^{2i\phi})+e^{i\phi}(1-e^{2Mi\phi})(1-e^{(N-M)i\phi}-e^{(N-M+1)i\phi})|^2}{|(M-1)(1-e^{2i\phi})-e^{2i\phi}(1-e^{2(M-1)i\phi})|^2+|M(1-e^{2i\phi})+e^{i\phi}(1-e^{2Mi\phi})(1-e^{(N-M)i\phi}-e^{(N-M+1)i\phi})|^2} &   (\mathrm{ii})
\end{cases}
\end{eqnarray}
which further yields the condition for the reciprocity $\sigma^G(\omega)=0$ in the general braided topology (i)
\begin{eqnarray}
N\sin\phi[\cos(N-1)\phi+\cos N\phi-\cos M\phi-\cos(M+1)\phi]+\sin N\phi[1+\cos\phi-\cos(M-N)\phi-\cos(M-N+1)\phi]=0\nonumber
\end{eqnarray}
and in the general braided topology (ii)
\begin{eqnarray}
M\sin\phi[\cos(M-1)\phi+\cos M\phi-\cos N\phi-\cos(N+1)\phi]+\sin M\phi[1+\cos\phi-\cos(N-M)\phi-\cos(N-M+1)\phi]=0.\nonumber
\end{eqnarray}
\end{widetext}

\begin{figure} [hbt]
\centering
\includegraphics[width=0.46\textwidth]{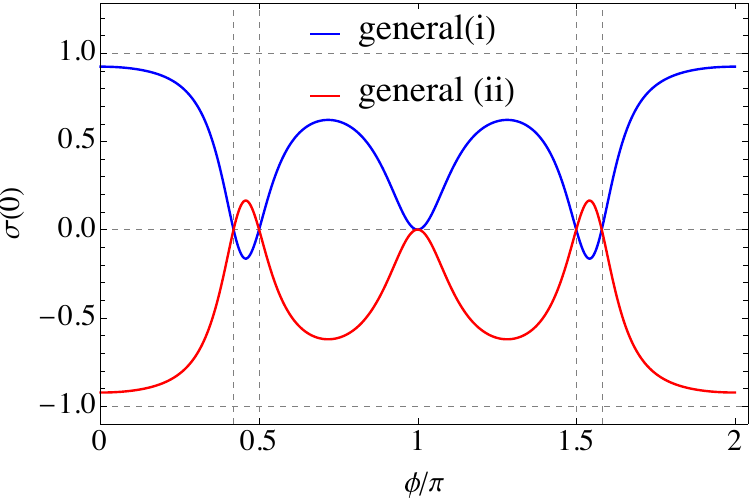}
\caption{(Color online) The numerical simulations of the nonreciprocal strength $\sigma(0)$ in the general braided topologies. The gridelines indicate the positions $\sigma(0)=\{-1,0,1\}$. The coupling numbers are set as $M=3, N=2$ and $N=3, M=2$ to plot (i) and (ii), respecitively. Compared to Fig. \ref{SigmaFig1}, the additional phase $e^{2i\phi}+e^{4i\phi}$ contributed by the separated part also participates in the destructive interference, resulting in extra reciprocal points $\phi\simeq\{0.41875\pi, \pi, 1.58125\pi\}$. }
\label{SigmaFig2}
\end{figure}

\subsection{Numerical simulations at the number of coupling points $M=3, N=2$ and  $N=3, M=2$ }
To compare with strict braided topologies, we set  $M=3, N=2$ and $N=3, M=2$ to consider the general braided topologies (i) and (ii), respectively. We also assume that the system is resonantly driven to a steady state with $\Delta_1= \Delta_2=\omega =0$. Under the above assumptions, the nonreciprocal strength $\sigma(0)$ takes the form
\begin{equation}
\sigma(0)=
\begin{cases}
\frac{|2e^{i\phi}+e^{2i\phi}+e^{3i\phi}+e^{4i\phi}|^2-1}{|2e^{i\phi}+e^{2i\phi}+e^{3i\phi}+e^{4i\phi}|^2+1} & general~(\mathrm{i})\\
\frac{1-|2e^{i\phi}+e^{2i\phi}+e^{3i\phi}+e^{4i\phi}|^2}{1+|2e^{i\phi}+e^{2i\phi}+e^{3i\phi}+e^{4i\phi}|^2} & general~(\mathrm{ii})
\end{cases}.\label{GGeneralBraidedSigma}
\end{equation}
Compared to Eqs. \eqref{GSigmaA} and \eqref{GSigmaB}, the additional separated component in the general braided topologies induces a phase $e^{2i\phi}+e^{4i\phi}$. This phase also contributes to destructive interference, resulting in additional reciprocal points $\phi\simeq\{0.41875\pi, \pi, 1.58125\pi\}$, as shown in Fig. \ref{SigmaFig2}.

\begin{figure*} [bt]
\centering
\includegraphics[width=0.92\textwidth]{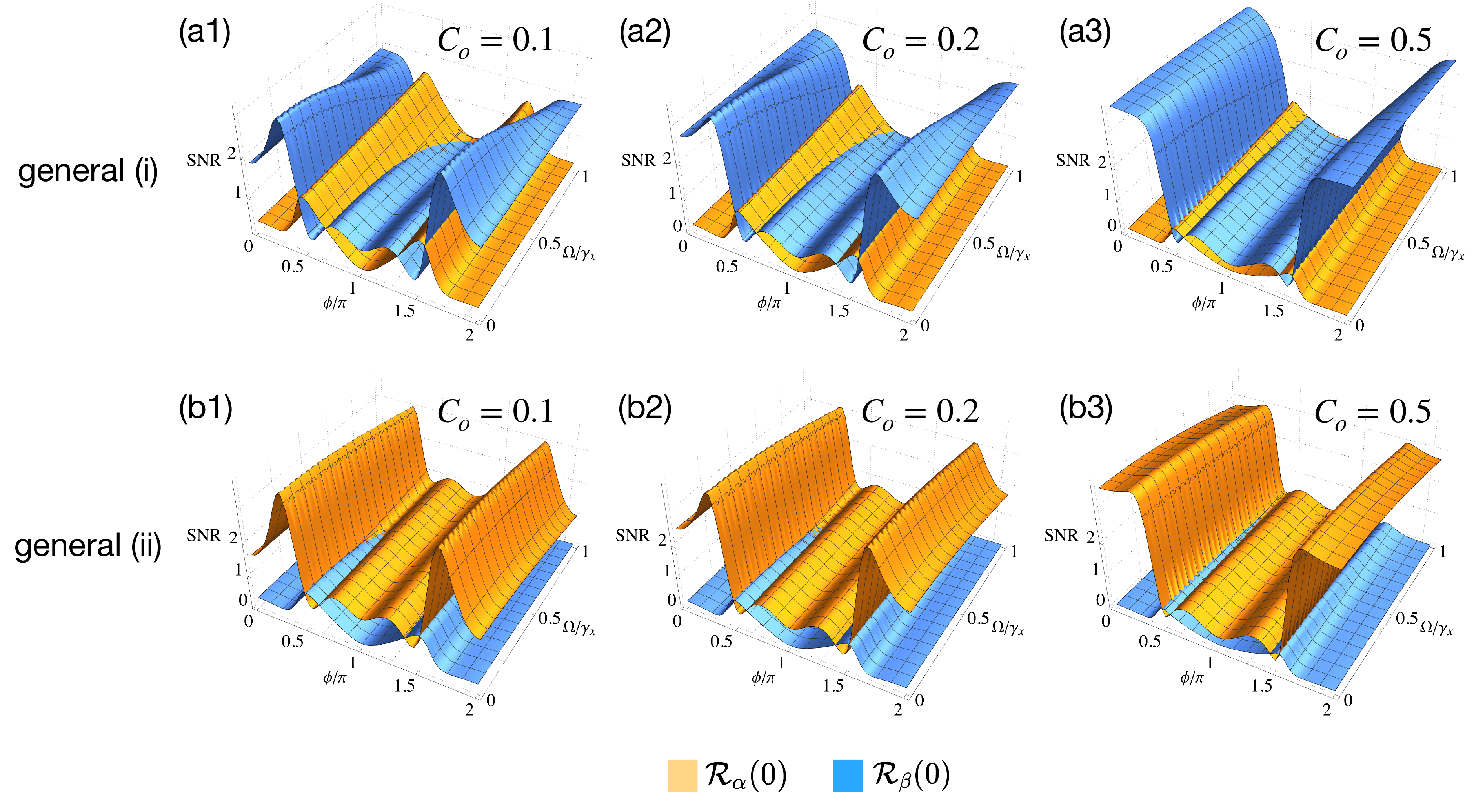}
\caption{(Color online) The numerical simulations of the SNRs $\mathcal{R}_{\alpha} (0)$ and $\mathcal{R}_{\beta} (0)$ in the general braided topologies. Parameters used are consistent with those in Fig. \ref{SNR1}. Compared to Fig. \ref{SNR1}, the phase introduced by the additional separated components enhances both SNRs, $\mathcal{R}_{\alpha} (0)$ and $\mathcal{R}_{\beta} (0)$ outside the interval $\phi \simeq [0.41875\pi, 1.58125\pi]$, while decreasing them within this interval. Aside from this, the other results are similar to those in Fig. \ref{SNR1}. }
\label{SNR2}
\end{figure*}

\begin{figure*} [bt]
\centering
\includegraphics[width=0.92\textwidth]{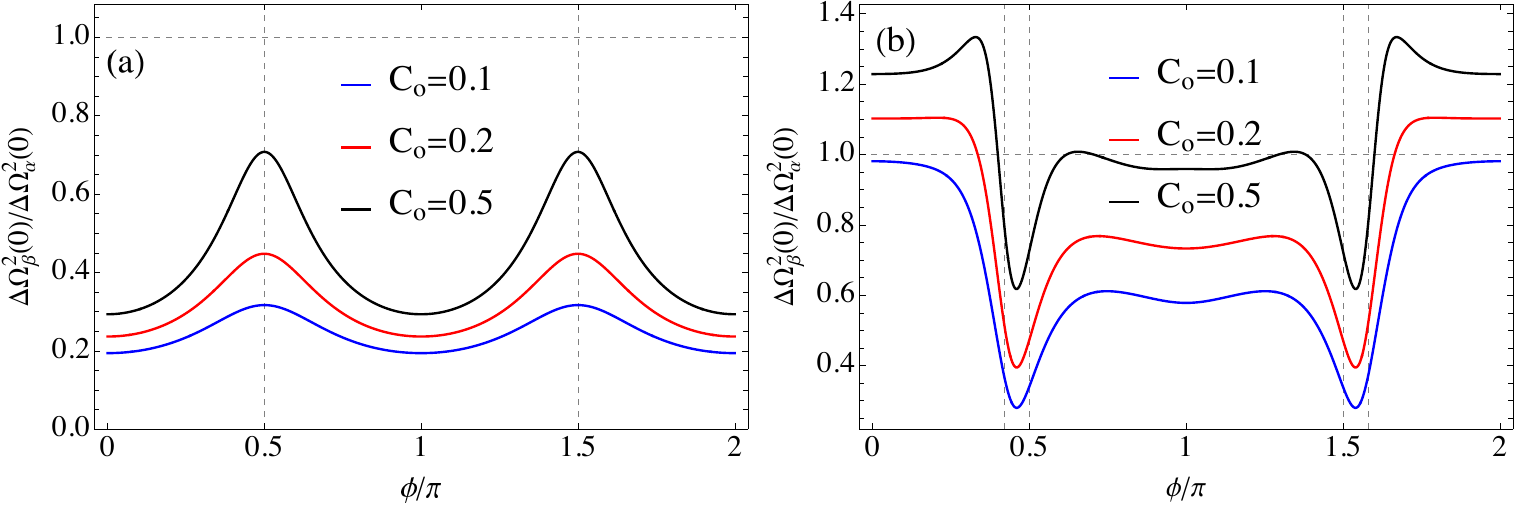}
\caption{(Color online) The numerical simulations of the sensitivity ratio $\Delta\Omega_\beta^2(0)/\Delta\Omega_\alpha^2(0)$ in the general braided topologies. In (a), the sensitivity ratio $\Delta\Omega_\beta^2(0)/\Delta\Omega_\alpha^2(0) $ exhibits a similar trend to that shown in Fig. \ref{Sensitivity1} (a), despite the introduction of additional separated components that contribute an extra phase. In (b), the sensitivity ratio $\Delta\Omega_\beta^2(0)/\Delta\Omega_\alpha^2(0)$ exceeds 1 outside the reciprocal points $\phi\simeq\{0.41875\pi,1.58125\pi\}$ due to this extra phase, even though the system remains constrained within the weak coupling regime. This observation indicates that the output signal $\mathcal{S}_\alpha(0)$ benefits from left nonreciprocity, resulting in increased sensitivity to changes in angular velocity $\Omega$. }
\label{Sensitivity2}
\end{figure*}

Also, the SNRs $\mathcal{R}_\alpha(0)$ and $\mathcal{R}_\beta(0)$ read
\begin{subequations}
\begin{eqnarray}
\mathcal{R}_\alpha(0)=
\begin{cases}
2|1-\frac{f^g_1\kappa+\gamma_x+\frac{4\Omega^2}{\gamma_y}}{F^g_1\frac{\kappa}{2}+F^g_2(\frac{\gamma_x}{2}+\frac{2\Omega^2}{\gamma_y})}|^2 & general~\mathrm{(i)}\\
2|1-\frac{f^g_2\kappa+\gamma_x+\frac{4\Omega^2}{\gamma_y}}{F^g_1\frac{\kappa}{2}+F^g_2(\frac{\gamma_x}{2}+\frac{2\Omega^2}{\gamma_y})}|^2 & general~\mathrm{(ii)}
\end{cases}\label{RAGeneral}\\
\mathcal{R}_\beta(0)=
\begin{cases}
2|1-\frac{f^g_2\kappa}{F^g_1\frac{\kappa}{2}+F^g_2(\frac{\gamma_x}{2}+\frac{2\Omega^2}{\gamma_y})}|^2 & general~\mathrm{(i)}\\
2|1-\frac{f^g_1\kappa}{F^g_1\frac{\kappa}{2}+F^g_2(\frac{\gamma_x}{2}+\frac{2\Omega^2}{\gamma_y})}|^2 & general~\mathrm{(ii)}
\end{cases}\label{RBGeneral}
\end{eqnarray}
\end{subequations}
where $f^g_1=1+\sqrt{C_o}(\frac{3}{2}+e^{2i\phi}+e^{3i\phi})$,
 $f^g_2=1+\sqrt{C_o}(1-2e^{i\phi}-e^{3i\phi}-e^{4i\phi})$, $F^g_1=1+C_o(\frac{3}{2}+e^{i\phi}+\frac{1}{2}e^{2i\phi}+e^{3i\phi})+\sqrt{C_o}(\frac{5}{2}+e^{i\phi}+2e^{2i\phi}+e^{3i\phi})$, and $F^g_2=F^s_2=1+\sqrt{C_o}(1+e^{2i\phi})$. Compared to the strict braided topologies, the phase introduced by the additional separated components enhances both SNRs outside the interval $\phi\simeq[0.41875\pi,1.58125\pi]$ but decreases them within this interval, as shown in Fig. \ref{SNR2}. Aside from this, the other results are consistent with those presented in Fig. \ref{SNR1}.

Correspondingly, the sensitivity are given by
\begin{subequations}
\begin{eqnarray}
&&\Delta\Omega_\alpha^2(0)\simeq
\begin{cases}
\frac{\gamma_y\kappa}{16|\alpha|}|\frac{(F^g_1)^2}{f^g_2\sqrt{C_o}}| & general~\mathrm{(i)}\\
\frac{\gamma_y\kappa}{16|\alpha|}|\frac{(F^g_1)^2}{F^g_1-F^g_2f^g_2}| & general~\mathrm{(ii)}
\end{cases}\label{SensitivityAG}\\
&&\Delta\Omega_\beta^2(0)\simeq
\begin{cases}
\frac{\gamma_y\kappa}{16|\alpha|}|\frac{(F^g_1)^2}{f^g_2F^g_2}| & general~\mathrm{(i)}\\
\frac{\gamma_y\kappa}{16|\alpha|}|\frac{(F^g_1)^2}{f^g_1F^g_2}| & general~\mathrm{(ii)}
\end{cases}\label{SensitivityBG}
\end{eqnarray}
\end{subequations}
as plotted in Fig. \ref{Sensitivity2}. As shown in Fig. \ref{Sensitivity2} (a), the sensitivity ratio$\Delta\Omega_\beta^2(0)/\Delta\Omega_\alpha^2(0)$ remains consistent with that shown in Fig. \ref{Sensitivity1} (a), despite the introduction of additional separated components that add an extra phase. This extra phase does not influence the sensitivity ratio in the general braided topology (i). When comparing Fig. \ref{Sensitivity2} (b) with Fig. \ref{Sensitivity1} (b), significant differences emerge outside the reciprocal points $\phi\simeq\{0.41875\pi,1.58125\pi\}$, where the sensitivity ratio $\Delta\Omega_\beta^2(0)/\Delta\Omega_\alpha^2(0)$ in the general braided topology (ii) exceeds 1, even though the system remains within the weak coupling regime. This finding suggests that left nonreciprocity enhances the output signal $\mathcal{S}_\alpha(0)$, making it more sensitive to changes in angular velocity $\Omega$.

Based on previous results, one can infer that increasing the number of coupling points leads to more complex interference and alters reciprocal points. Consequently, this can enhance the output SNR to a certain extent; however, the impact on sensitivity must be analyzed specifically in relation to the number of specific coupling points.

\bibliography{GiantGyroscope-1024}

\end{document}